\documentclass[review]{elsarticle}
\graphicspath{ {./figures/} }
\usepackage{hyperref}
\usepackage{float}
\usepackage{verbatim} %comments
\usepackage{apalike}
\restylefloat{figure}
\restylefloat{table}
\usepackage{amssymb}
\usepackage{epigraph}

\usepackage{fullpage}
\usepackage{multicol}
\usepackage{subcaption}
\usepackage{graphicx}
\usepackage{todonotes}

\usepackage{algpseudocode}
\usepackage{algorithm}
\usepackage{ragged2e}

\makeatletter
\def\ps@pprintTitle{%
 \let\@oddhead\@empty
 \let\@evenhead\@empty
 \def\@oddfoot{\centerline{\thepage}}%
 \let\@evenfoot\@oddfoot}
\makeatother

\providecommand{\keywords}[1]
{
  \small	
  \textbf{\textit{Keywords---}} #1
}

\usepackage{color}
\algdef{SE}[VARIABLES]{Variables}{EndVariables}
   {\algorithmicvariables}
   {\algorithmicend\ \algorithmicvariables}
\algnewcommand{\algorithmicvariables}{\textbf{Inputs}}

\journal{Expert Systems with Applications}

\biboptions{authoryear}

\begin{document}
\begin{frontmatter}

\title{Proactive Query Expansion for Streaming Data Using External Sources}
         
\author[label1]{Farah Alshanik }
\ead{falshan@g.clemson.edu}
\author[label2]{Amy Apon}
\ead{aapon@clemson.edu}
\author[label3] {Yuheng Du}
\ead{yuheng.du.hust@gmail.com}
\author[label4]{Alexander Herzog }
\ead{aherzog@clemson.edu}
\author[label5]{Ilya Safro }
\ead{isafro@udel.edu}

%\cortext[cor1]{}
\address[label1]{ School of Computing, Clemson University, 100 McAdams Hall,
Clemson, 29634, SC, US }
\address[label2]{ School of Computing, Clemson University, 100 McAdams Hall,
Clemson, 29634, SC, US}
\address[label3]{ School of Computing, Clemson University, 100 McAdams Hall,
Clemson, 29634, SC, US}
\address[label4]{ School of Computing, Clemson University, 100 McAdams Hall,
Clemson, 29634, SC, US}
\address[label5]{Computer and Information Sciences , University of Delaware, 430 Smith Hall, 
Newark, 19716, DE, US}

\begin{abstract}
Query expansion is the process of reformulating the original query by adding relevant words. Choosing which terms to add in order to improve the performance of the query expansion methods or to enhance the quality of the retrieved results is an important aspect of any information retrieval system. Adding words that can positively impact the quality of the search query or are informative enough play an important role in returning or gathering relevant documents that cover a certain topic can result in improving the efficiency of the information retrieval system. Typically, query expansion techniques are used to add or substitute words to a given search query to collect relevant data. In this paper, we design and implement a pipeline of automated query expansion. We outline several tools using different methods to expand the query. Our methods depend on targeting emergent events in streaming data over time and finding the hidden topics from targeted documents using probabilistic topic models. We employ Dynamic Eigenvector Centrality to trigger the emergent events, and the Latent Dirichlet Allocation to discover the topics. Also, we use an external data source as a secondary stream to supplement the primary stream with relevant words and  expand the query using the words from both primary and secondary streams. An experimental study is performed on Twitter data (primary stream) related to the events that happened during protests in Baltimore in 2015. 
The quality of the retrieved results was measured using a quality indicator of the streaming data: tweets count, hashtag count, and hashtag clustering. 
Results indicate that adding words from the secondary stream can significantly improve the quality of search queries and return more relevant information that covers a certain topic. Another experimental study is performed to measure the effectiveness of the proposed methods to predict future emerging events or to predict the conversations from previous time intervals using the precision of a different set of hashtags. Results show that our method increases the precision. 

Reproducibility: code and results are available
at \url{https://github.com/FarahAlshanik/QE}

%In this paper, we propose a method called the historical query expansion which applies  a smaller discrete time interval to capture the dynamic features in textual data. In addition, due to As the nature of streaming data, Twitter stream contains emerging events that are constantly changing and therefore not predictable using static queries a  Thus, we also adopt DEC to find the time intervals that have the emergent events to expand the query and tracking on conversational content using hashtags. In this study, we use the Baltimore Tweets  as an example to capture the emergent events, expand the query and predict the conversational content.

\end{abstract}

\end{frontmatter}


%\newpage
%\textbf{\color{red} \Large TOC for internal use only! Will be deleted before submitting the paper!}
%\tableofcontents
%\newpage

\section{Introduction}
\label{introduction}
Social media streaming data (e.g., Twitter messages or Facebook posts) have become a primary source to analyze public opinion \citep{d2019monitoring}, track user sentiment \citep{giachanou2016like, martinez2014sentiment}, or  study emergent safety events \cite{avudaiappan2017detecting} including public health crises \citep{guidry2017ebola,glowacki2016identifying,bolotova2017detecting}, natural disasters \citep{roy2010impact, david2016tweeting, ullah2021rweetminer, DBLP:conf/ssd/HubigFZYG17}, and political or social movements \citep{vaccari2015political,keib2018important}. Queries used to draw data from these high-volume, high-velocity, real-time sources typically require a set of words to filter the data. For example, tracking the public sentiment surrounding the COVID-19 pandemic on Twitter may rely on words such as ``covid'', ``corona'', and ``lockdown'' to filter out relevant messages. Such queries initiated by a static word list can be problematic because they reflect the domain expertise of users, and therefore reflect their biases or jargon, which can result in the exclusion of words required to retrieve relevant information. Static words also fail to keep up with changes in language and emergent words which results in incomplete data.

Information retrieval systems typically use query expansion techniques to enhance the initial user query, e.g., by adding inflected forms, cognates, and related words manually retrieved from the text. \citep{schutze1997cooccurrence,zhang2009concept}. We propose a novel query expansion technique that addresses the challenges of analyzing data from high-volume, high-velocity social media streams. We argue that effectively filtering a data stream in an environment in which language and terms can rapidly change does not necessarily rely only on information from the stream itself. Instead, we develop a query expansion technique that integrates words from the current stream with \emph{external data sources} (in our experiments, newspaper archives) in order to predict the occurrence of relevant words that have not appeared in the stream yet. This algorithm enables the construction of new queries that effectively capture emergent events that a user may not have anticipated when initiating the data collection stream. The purpose of using the \emph{external data sources} appears when we have a stream of data and we want to predict something that has not happened yet instead of using only the stream that is limited to the available information at a specific time.  The external data source will enrich queries with words that cover future topics. For example, using the static word "protest" to filter stream at a specific time $x$ will fail to retrieve the related information of the word "protest". Generally, the “protest” results in violence, looting, and sometimes in curfew but these words or "events" will appear in the future as a result of protest and do not appear in the stream yet. In order to retrieve the information that correlated with protest, we need to augment the query with information from an external source. In another example, COVID-19 has impacted the lives of millions of people across the globe in different ways. Travel restrictions and business closures are some of the events that happened in response to the COVID-19 pandemic. Predicting such events from a stream of data at a specific time using static query requires an external source with information related to similar event happened in the past to enrich the query with correlated information. For example, if we have an external source (news articles) related to the MERS coronavirus that has appeared in news since 2012, and we want to filter a stream (tweets about COVID-19) at a specific time $x$ using the static query ``coronavirus''. The MERS coronavirus appeared in news along with some business closures and travel restrictions. Relying only on the information from the stream for query expansion means that the static word “coronavirus” might be unable to predict the events that may appear in the future at time $x+n$ as a result of the COVID-19 pandemic. Using the external source about the MERS coronavirus allows our system to search for old words that were related to ``coronavirus'' to augment our static query  ``coronavirus" with future events or topics such as travel restrictions and business closures. Our system should be able to predict the related information of the static word based on the available external source and augment the static query with dynamic words that may appear in the future and have a correlation with a static query in such a way.

We demonstrate the validity and effectiveness of our approach with an analysis of Twitter messages surrounding the 2015 Baltimore protests. Our data consists of more than 20.5 million tweets collected over two weeks. We use this data to evaluate the performance of our query expansion method against two alternative approaches. the first approach expands user queries with words extracted from a probabilistic topic model of the stream. The second approach reinforces them with emergent words extracted from the stream. We conduct several experiments in which we evaluate the performance of our query expansion along three metrics applied to the retrieved results: volume (measured by tweet count), relevance (measured by hashtag count), and conciseness (measured through hashtags clustering). We find that our methods outperform alternative approaches, exhibiting particularly good results in identifying future emergent topics.

The remainder of our paper is organized as follows: Section 2 covers related work. The system overview of our proactive query expansion method and three methods  for expanding query in addition to the reference model are described in Section 3. Algorithm is described in Section 4. Evaluation and Results are provided in Section 5, and Section 6 respectively and section 7 concludes the paper.

\section{Related Work}
\label{Related Work}
Query expansion has been an active area of research and several studies have sought an automated method to deal with the word mismatch in information retrieval. \citep{xu2017quary} perform automatic query expansion using three representative techniques. The first technique is the global analysis based on the method introduced by \citep{jing1994association}. The global analysis technique creates a thesaurus-like database with a ranked list of phrases for a given query. The method is known as the global analysis approach because the association database it uses considers the entire collection of documents, and the process is frequently computationally intensive.
%Using the global analysis technique a ranked list of words is generated using a thesaurus-like database for a given query. The database takes into account \added[id=is]{what do you mean when you say that db takes into account?} all the documents in the corpus, which makes the process computationally heavy.
The task in \cite{xu2017quary} is different from our task in which we use the streaming data as a primary stream to the query. This means the thesaurus-like database used in \cite{jing1994association} is not directly applicable. Besides, the database in this solution requires to be updated with each new tweet, which makes the method fail in large-scale data \citep{du2019streaming}. 

The second approach introduced by \citep{xu2017quary} is the local feedback method, which overcomes the drawback of the global analysis by using the documents in the query results to generate a list of top-ranked words instead of using the entire corpus. The efficacy of this method crucially depends on the quality of the query result itself. The reliability of the local feedback method, therefore, remains an issue even it is less expensive to perform \citep{du2019streaming}. Local context analysis is the third technique introduced by \citep{xu2017quary}, which uses a combination of the global analysis approach and the local feedback approach. It uses the ranked query results to identify the top concepts (noun groups share the same semantic meaning). Based on the distance of each concept to the original query in the global thesaurus and their TF-IDF scores, the local context analysis picks new words. This method achieves better performance than using either global analysis or local feedback separately. However, it also requires a static metric to rank the documents in the query result. 

For querying streaming social data, the metrics to evaluate the goodness of the query results are often dynamically changing and may comprise a mixture of various sub-metrics \citep{du2019streaming}. Hence, it is not feasible to directly use the local context analysis method introduced by \citet{xu2017quary}. The query expansion method proposed in \citet{massoudi2011incorporating} uses tweet data as the query platform. \citet{massoudi2011incorporating}'s approach is similar to our query expansion work in that they employ a time-based indicator to deal with the data streaming’s dynamic nature, which is similar to our method of measuring query quality using a dynamic metric. \citet{massoudi2011incorporating} use repost count and followers of posts as an indicator of a tweet’s quality which changes with time. Our approach uses tweet count and hashtags information as a quality indicator of query results.

Many studies use different techniques to detect the emergent events in the streaming data \citep{saeed2019enhanced,chen2018social,adedoyin2016rule,avudaiappan2017detecting,chen2021multi}. The application of topic models has been an active area of research in informational retrieval, and several studies have sought an automated method for using topic models in query expansion. \citep{wei2006lda} apply Latent Dirichlet Allocation (LDA)   to improve the retrieval results using cluster-based models. \citep{yi2008evaluating} use Mixture of Unigrams \citep{xu1999cluster}, LDA \citep{blei2003latent}, and the Pachinko Allocation Model \citep{li2006pachinko} to integrate the topic models into the retrieval process. \citep{zhou2009latent} use topic models for re-ranking initial retrieval results. \citep{ye2011finding} use LDA in query expansion to improve the performance of relevance feedback. \citep{christidis2012using} use topic modeling to enhance the search and recommendation functionalities of Enterprise Social Software. \citep{serizawa2013study} expand the initial query by using the latent topic information on the documents retrieved at the first search. Similar to this line of work, we propose a method that uses topic distributions of the targeted documents in addition to an external data source to expand the query. To the best of our knowledge, there is no existing study that uses topic modeling with external sources to expand the initial query in streaming data.

% \section{Background}
% AH: This should be added to the Related Works section or described below when we discuss the experimental design

% The low-dimensional representation model, fastText, is utilizing the skip-gram model. representation \citep{bojanowski2017enriching}. The fastText maps words into a low-dimensional space and uses subword information  revealing non-trivial context based relationships between them. Many syntactic and semantic relationships between words can be defined by using simple algebraic operations on the word vectors.

% Dynamic Eigenvector Centrality (DEC) \citep{avudaiappan2017detecting} , is graph based technique. It extracts the emergent words from a document stream based on dynamic semantic graphs. DEC, in its simplest form, construct a network in which the nodes and undirected edges correspond to words and co-occurrences of words in a stream of documents, respectively. DEC use the constructed network to rank and extract top-ranked emergent words.

% Latent Dirichlet Allocation (LDA), a generative probabilistic model, is an unsupervised learning method used for extracting latent topics from a large set of documents. The LDA is uses a three-level Bayesian model to fit the generative process, it represents documents as random mixtures over latent topics and represents each topic as a distribution over words \citep{blei2003latent}.

%\section{Proposed Pipeline}
\section{Proactive Query Expansion Method}
\label{sec:algorithm}

%\added[id=is]{I don't really like this wording ``historical query expansion''. It creates an impression that we are dealing with old information retrieval. How about ``proactive query expansion''? It is more about the future. If you agree, let's change everywhere.} 

\subsection{System Overview}   
\begin{figure}[tbp]
   \includegraphics[width=\textwidth]{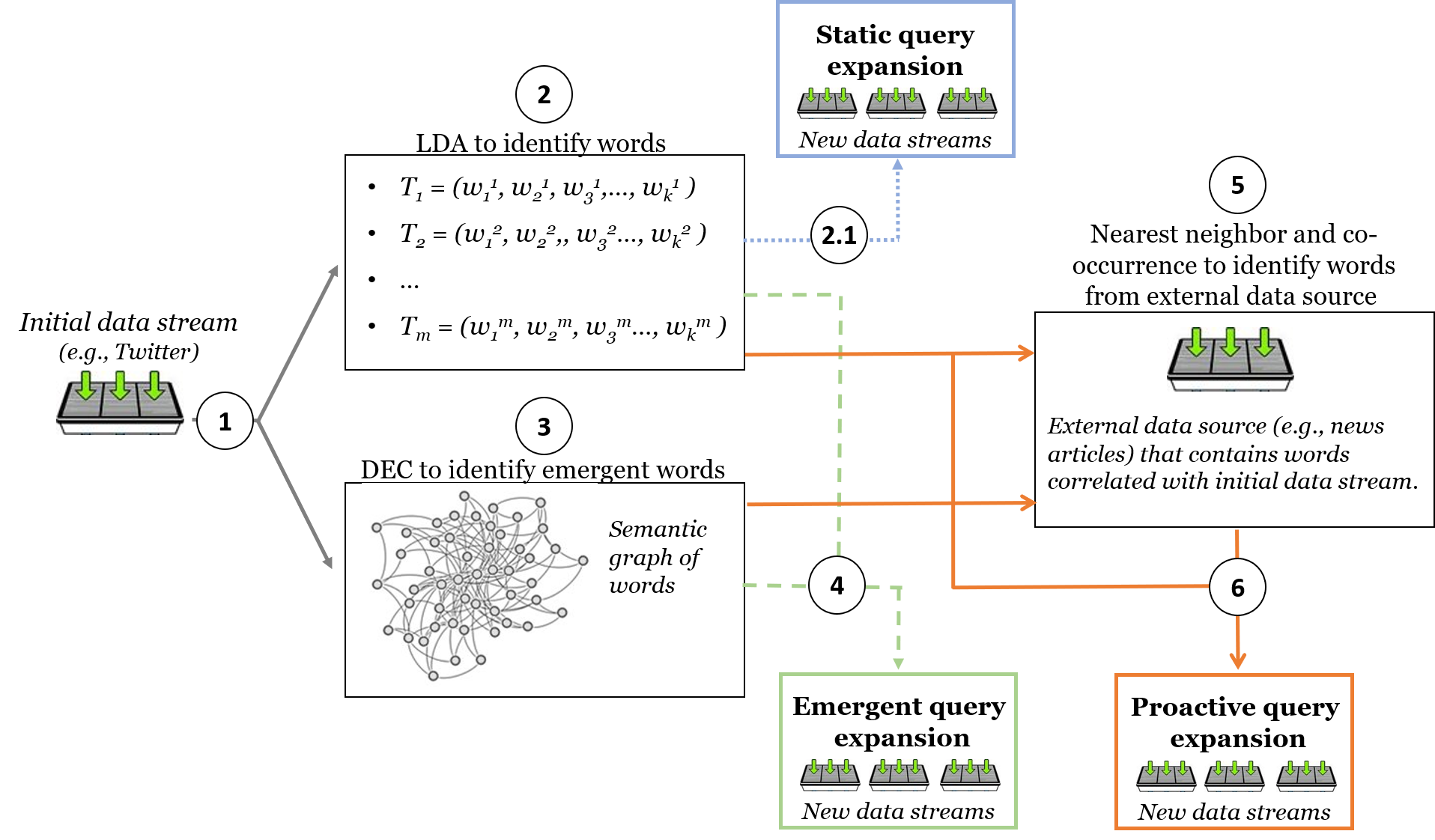}
   \caption{Simplify illustration of the core logic of how queries constructed}
\vspace*{2em}    
\begin{tabular}{lp{0.85\textwidth}}
\vspace*{0.5em} 
\textbf{Step 1.} & Detect emergent events in the initial stream using DEC
\cite{avudaiappan2017detecting} to trigger query expansion process.\\ 
\vspace*{0.5em} 
\textbf{Step 2.} & Use LDA to identify words in the initial stream.\\
\vspace*{0.5em} 
\textbf{Step 2.1}& Use LDA words to construct \ \textbf{static} queries.
 \\
 \vspace*{0.5em} 
\textbf{Step 3.} & Use DEC to identify emergent words in the initial stream.
 \\
 \vspace*{0.5em} 
\textbf{Step 4.} & Combine LDA and DEC words to construct \textbf{emergent} queries. \\ 
\vspace*{0.5em} 
\textbf{Step 5.} & Use nearest neighbor and co-occurrence applied to external data source to identify words that are correlated with LDA/DEC words extracted from the initial stream. \\ 
\vspace*{0.5em} 
\textbf{Step 6.} & Combine LDA, DEC and words from external data source to construct \textbf{proactive} queries using (vector space or co-occurrence). \\ 

\hline
\end{tabular}   
   \label{fig:sys_overview}
   \end{figure}

We introduce the proactive query expansion approach to detect emerging events in streaming data. Our approach utilizes external data sources to expand and enrich user proactive queries with words that do not appear in the stream yet but are highly correlated with emergent words in the existing stream. By adding these proactive words, our system can construct queries that capture emergent events that were not anticipated when initiating the data collection stream. 
%\added[id=is]{In this paragraph you need to inject notation from the figure and continue using the same notation in the algorithms. e.g., $\textsf{S}_1$ and $\textsf{S}_2$ already denote streams.}
In Fig.~\ref{fig:sys_overview} we illustrate our proactive query expansion system. First, the initial data stream (in our application it is Twitter) is being monitored for emergent events to trigger query expansion process using the method introduced by \cite{avudaiappan2017detecting} who use Dynamic Eigenvector Centrality to detect emergent words. 
If an emergent event is detected, a new set of queries is triggered in step 2 by collecting relevant words from the initial data stream  using LDA. In step 3, we use DEC to identify the emergent words in the initial stream and combine these words with LDA words in step 4 to construct new queries. In step 5, we identify  proactive words from an external data source. These words are determined by using two methods that in the past were correlated with words extracted from the initial data stream. In the final step (step 6), in addition to the LDA words and DEC words each query is expanded with proactive words from the external data source.

\subsection{Query Expansion Methods}   
At the core of our system is the proactive query expansion method, which combines words retrieved from the primary stream with novel words extracted from the external source (archival news in our application) with the goal to anticipate words that have not appeared yet. Our method belongs to the class  of approaches that employ LDA to build new queries.

%{Our method is an extension of existing approaches, namely using LDA to build new queries.} \deleted[id=is]{In this section, we first explain how existing research uses LDA to build new queries and then discuss how our method builds on these approaches. }

A large stream of literature uses LDA for query expansion \cite{christidis2012using, serizawa2013study}. LDA estimates latent topics from a given corpus, where each topic represents a ranked list of the words included in the corpus. Overall, using LDA for query expansion means that each topic discovered by LDA serves as a new query restricted to the top-ranked words in each topic. words in LDA are ranked by their relative importance for each topic. However, when applied to dynamic data, this approach ignores emergent words which is a significant disadvantage. A key characteristic of high-volume, high-velocity streaming data, such as Twitter, is that topics can change rapidly. Relying on LDA alone for query expansion means that the extracted words might be unable to capture emergent topics. 

Several studies propose different techniques to detect emergent events in streaming data \cite{saha2012learning,benhardus2013streaming, adedoyin2016rule,avudaiappan2017detecting}. In our system, we use the Dynamic Eigenvector Centrality (DEC) method to detect emergent words  \cite{avudaiappan2017detecting} because of its ability to detect meaningful, less noisy, and more
interpretable information in data streams than frequency-based measures used in  \cite{saha2012learning,benhardus2013streaming}. A natural improvement of LDA in the context of streaming data is therefore to combine words extracted with LDA with words identified as emergent based on the DEC method.

Combining LDA with DEC to construct new queries (LDA-DEC) overcomes the static nature of LDA and, we argue, is better suited to construct queries for streaming data where topics are  rapidly changing. However, this method still only relies on words extracted from the current stream, which means that the resulting queries will potentially miss words that have not appeared in the stream yet. For this reason we propose the proactive query expansion  which extends the LDA-DEC approach by adding words that, historically, were correlated with the words identified from the stream. We detect these correlated words through two different methods, both of them rely on the same external data source. In our first method, we construct a low-dimensional representation of the external data. We then use this vector space of proactive words to identify words that are close to those detected by the combined LDA-DEC method, using nearest neighbor search as proximity measure. In our second approach, we select proactive words based on their co-occurrence frequency with the LDA and DEC words. 

%\todo{is: why 4? the static is not new, it is only for reference}

The above discussion leads to three alternative query expansion methods in addition to the reference method for streaming data, which are summarized in Table~\ref{tab:expansion_methods2}. Method 1, Static, only uses the information from the current stream, which we include here as a benchmark case against which we will evaluate the other methods. Method 2, Emergent, combines the LDA words with the emergent words identified with DEC. Method 3, Proactive using vector space (VS), is our first proactive query expansion that adds words from the vector space constructed from the external data source. Method 4, Proactive using co-occurrence, is our second proactive query expansion that adds  words from the external data source based on their co-occurrence frequency.

%\begin{table}[htbp]
%\vspace{-.02in}
%\caption{Summary of Proposed Query Expansion Methods }\label{tab:expansion_methods}
%\begin{center}
%\begin{tabular}{lccc}
%\hline
%& \multicolumn{3}{c}{\textbf{Type of Information Source}}\\
%& \textbf{Static} & \textbf{Emergent} & \textbf{Historical}\\
%\hline
%Method 1: LDA & \checkmark & - & -\\[0.5em]
%Method 2: LDA \& DEC & \checkmark & \checkmark & -\\[0.5em]
%Method 3: LDA \& DEC \& Vector Space & \checkmark & \checkmark & \checkmark\\[0.5em]
%Method 4: LDA \& DEC \& Co-Occurrence Frequency & \checkmark & \checkmark & \checkmark\\
%\hline
%\end{tabular}
%\end{center}
%\vspace{-.1in}
%\end{table}

\begin{table}[htbp]
\vspace{-.02in}
\caption{Summary of Proposed Query Expansion Methods }\label{tab:expansion_methods }\label{tab:expansion_methods2}
\begin{center}
\begin{tabular}{lcccc}
\hline
  & \multicolumn{4}{c}{\textbf{Method for word Identification}}\\\cline{2-5}
  & & & External & External  \\
  & LDA & DEC & Vector Space (VS) & Co-Occurrence (CO) \\
\hline
\vspace*{0.5em} 
Method 1 -- Static       & \checkmark & - & - & - \\
\vspace*{0.5em} 
Method 2 -- Emergent     & \checkmark & \checkmark & - & - \\
\vspace*{0.5em} 
Method 3 -- proactive VS & \checkmark & \checkmark & \checkmark & - \\
\vspace*{0.5em} 
Method 4 -- proactive CO & \checkmark & \checkmark & - & \checkmark \\
\hline
\end{tabular}
\end{center}
\vspace{-.1in}
\end{table}

\section{Algorithm }
%\todo{ ``Implementation'' is not a g/ood title here.}
In this section we introduce the proactive query expansion and give a detailed description of each query expansion method. We also compare the proactive query expansion method to the reference method (labeled "static") and the emergent method ("emergent"). In all query expansion methods, we define the algorithms (Algorithms~{1} to {4}) with reference to using data from Twitter as an example of streaming data. Table \ref{notation} summarizes the notations that will appear in the algorithms.

\begin{table}[htbp]\caption{Table of Notation }
\begin{center}
\label{notation}
\begin{tabular}{r c p{10cm} }
\hline
$S_{1}$& $:$ & Primary tweet stream\\
$S_{2}$& $:$ &External data source\\
%$S_{s}$& $:$ &Static tweet stream\\
$T$& $:$ & A set of topics result from LDA\\
$J_{s}$ & $:$ & Jaccard similarity \\
$th$ & $:$ &Specific threshold  \\
$w$ & $:$ &Time interval (window)  \\
$n$ & $:$ &Number of windows  \\
$d$ & $:$ &Number of top-ranked DEC words    \\
$s$ & $:$ & A tweet in the primary tweet stream\\
$t$ & $:$ & A given topic in $T$\\ 
$m$ & $:$ & Number of LDA topics\\ 
$k$ & $:$ & Number of top-ranked LDA words\\ 
$l$ & $:$ & Query length\\
$Q$ & $:$ & Query result for all topics\\
$q_{t}$ & $:$ & Query result associated with a given topic $t$\\ 
%$DEC_{w}$ & $:$ & Top ranked DEC words that do not appear in the top LDA words\\
$V$ & $:$ & External semantic vector space \\
%$nearest$ & $:$ & a set of neighboring words of specific word $w_{t}$ \\
$F$ & $:$ & External bi-grams dictionary \\
%$freq$ & $:$ & set of highest frequency words of specific word $w_{t}$ \\
$D$ & $:$ & DEC words for time interval $w$\\
$i$ & $:$ & Number of nearest neighbor words to return for a specific word \\
$j$ & $:$ & Number of highest frequency words to return for a specific word \\
\hline
\end{tabular}
\end{center}
\label{tab:TableOfNotation}
\end{table}
We process a data stream by discretizing it into time intervals (called window), each of length $w$ units (in our implementation we use minutes). We use the DEC metric introduced in \cite{avudaiappan2017detecting} to extract the top-ranked emergent words. We trigger the query expansion if the set of  emergent words in the current window indicate that an emergent event is occurring. To this end, we calculate the Jaccard similarity ($J_{s}$) of the top $d$ DEC words between the current window and the $P$-previous windows. If the Jaccard similarity is less than or equal to some threshold $th$, we assume an emergent event is occurring that will require a new search query with new keywords not currently captured in the query that has initiated the current stream. At this point in the stream, we execute Steps 1--6 from Figure~\ref{fig:sys_overview} to construct the following three queries: 

\textbf{Static}: Query expansion using LDA words. LDA is used to generate $m$ topics from the current time windows. Each topic represents a ranked list of the words included in the current window which reveals a discussion theme for the topic. Using LDA for query expansion means using each topic estimated with LDA as a new query restricted to the top-ranked words in each topic. The algorithm procedure of this method is described in Algorithm 1. After generating a set of topics, we expand the query and return the results that satisfy each expanded query, such that for a given  document $s$ from  the primary stream $S_{1}$ and a topic $t$ from a set of topics $T$, if we can find any $l$ LDA words in document $s$, then we add $s$ into the query result ($q_{t}$). Finally, the aggregated  query results $Q$ for all topics will be returned by the end of the procedure.

\begin{algorithm}
\caption{Query Expansion Using LDA words}
\begin{algorithmic}[1]
\label{lda}
\Variables
\State $Q$, Query result for all topics
\State $T$, A set of topics result from LDA
\State $S_{1}$, Primary tweet stream
%\State $s$, A tweet in the primary tweet stream
%\State $t$, A given topic in $T$
%\State $q_{t}$, Query result associated with a given topic $t$

\EndVariables
\Procedure{Static}{$Q$,$T$,$S_{1}$}       
    \State $Q$ $\leftarrow$ $\emptyset$ 
     \For{$t$ in $T$} \Comment{for each topic $t$ return the query result associated with it }
        \State $q_{t} \leftarrow \emptyset$  \Comment{initialize query result associated with topic $t$}
        \For{$s$ in $S_{1}$} \Comment{for each tweet $s$ in the primary stream $S_{1}$ check if the tweet has the LDA words}
         \State $cnt$=0 \Comment {count the number of LDA words in the tweets $s$ }
            \For{$v$ in $t$} \Comment {for each word in the topic $t$ check if the word $v$ in the tweet $s$}
                \If{$v$ in $s$} 
                    \State $cnt$+=1 \Comment{increase the number of LDA words in tweets $s$}
                \If{$cnt$ $\geq$ $l$}\Comment {if $cnt$ $\geq$ the length of the query $l$, add the tweet $s$ to the $q_{t}$ } 
                    \State $q_{t}$ $\leftarrow$	$q_{t}$ $\cup$ 	$s$ \Comment{ add the tweet $s$ to the query result $q_{t}$  }
                \EndIf 
            \EndIf 
            \EndFor  
        \EndFor  
         \State $Q$ $\leftarrow$	$Q$ $\cup$ \Comment{ query results for all topics}	$q_{t}$
    \EndFor
    \State return $Q$
\EndProcedure
\end{algorithmic}
\end{algorithm}

\textbf{Emergent}: Query expansion using LDA words and DEC words. We propose a query expansion method that combines words extracted with LDA with words identified as emergent based on the DEC method for a specific time window. Combining LDA with DEC to construct new queries overcomes the static nature of LDA and, we argue, is better suited to construct queries for streaming data where topics might rapidly change. For each topic $t$ and a time window $w$, we add $d$ top-ranked DEC words from $D$ that do not appear in the $k$ top-ranked LDA words for topic $t$ using function $Dec($w$,$t$,$D$,$d$)$ which saves the returned words in $Dec_{w}$. We used this condition to avoid redundant queries because we found that some top DEC words also appear in the top LDA words. By adding the DEC words to the topics, we will guarantee that the emerging topics are included in the query results. After adding the DEC words to each topic, we expand the query and return the results that satisfy each expanded query, such that for a given document $s$ from  the primary stream $S_{1}$, and a topic $t$ from a set of topics $T$, if we can find any $l$ LDA and DEC words in the document $s$, then we add document $s$ into query result $q_{t}$. Finally, the aggregated  query results $Q$ for all topics will be returned by the end of the procedure.

\begin{algorithm}
\caption{Query Expansion Using LDA words and DEC words}
\begin{algorithmic}[1]
\Variables
\State $Q$, Query result for all topics
\State $S_{1}$, Primary tweet stream
\State $T$, A set of topics result from LDA

%\State $s$, A tweet in the primary tweet stream
%%\State $T$, A set of topics result from LDA
%\State $t$, A given topic in $T$
\State $w$, Time interval
\State $D$, DEC words for time interval $w$
\State $d$, Number of top-ranked DEC words to return
%\State $DEC_{w}$, Top-ranked DEC words that do not appear in the top LDA words
%\State $q_{t}$, Query result associated with a given topic $t$
\EndVariables
\Procedure{Emergent}{$Q$,$T$,$w$,$S_{1}$,$d$,$D$}       
    \State $Q$ $\leftarrow$ $\emptyset$ 
     \For{$t$ in $T$}\Comment{for each topic $t$ return the query result associated with it }
        \State $q_{t} \leftarrow \emptyset$  \Comment{initialize query result associated with topic $t$}
         \State $DEC_{w} \leftarrow $DEC($w$,$t$,$d$ ,$D$)  \Comment{return $d$ top-ranked DEC words that do not appear in the top LDA words of topic $t$ for time interval $w$ }
         \State $t$ $\leftarrow$ $t$ $\cup$ 	$DEC_{w}$ \Comment{attach top $d$ DEC words to the topic $t$}
        \For{$s$ in $S_{1}$} \Comment{for each tweet $s$ in $S_{1}$ check if the tweet has the LDA and DEC words}
           \State $cnt$=0 \Comment {count the number of LDA and DEC words in the tweets $s$ }
             \For{$v$ in $t$} 
                \If{$v$ in $s$} \Comment {for each word in the topic $t$ check if the word $v$ in the tweet $s$}
                    \State $cnt$+=1 \Comment{increase the number of LDA and DEC words in tweets $s$ }
                \If{$cnt$ $\geq$ $l$} \Comment {if $cnt$ $\geq$ the length of the query $l$, add the tweet $s$ to the $q_{t}$ } 
                    \State $q_{t}$ $\leftarrow$	$q_{t}$ $\cup$ 	$s$\Comment{ add the tweet $s$ to the query result $q_{t}$  }
                \EndIf 
                \EndIf 
            \EndFor  
        \EndFor  
         \State $Q$ $\leftarrow$	$Q$ $\cup$ 	$q_{t}$
    \EndFor
    \State return $Q$
\EndProcedure
\end{algorithmic}
\end{algorithm}

\textbf{Proactive Vector Space}: Query expansion using LDA words, DEC words, and Vector Space. We extend the emergent query by using external data to add words that are correlated with the words identified from the stream but potentially haven't appeared in the initial stream yet. This method overcomes the limitation of the LDA-DEC method which only relies on words extracted from the current stream. Using this method allows us to capture future events or words that have not appeared in the stream yet. We used the fastText model \cite{bojanowski2017enriching} to generate a vector space $V$ to find the words’ nearest neighbor to each LDA and DEC word. The fastText model is utilized to construct an $n$-dimensional representation of each word in the external data called word embedding, each embedded word is represented as a vector of $n$ dimension. After representing each word by a vector, we used the fastText nearest neighbor method to find the closest words in space to a given word. The nearest neighbor method allows us to capture the semantic information of a given word. To find nearest neighbor words for a target word, this method computes the cosine similarity between the target word and all words in the vocabulary using the vector representation of the words. As an example of this process, Table~\ref{tab1} presents the top 10 nearest neighbor words for two words: "curfew" and "looting". These words represent key moments in the primary stream. As appear in the table, the top 3 nearest neighbor words for the word “curfew” are: “lockdown”, “nightfal”, and “impos” which means the static query “curfew” will be augmented by these words. For more investigation about the appearance of these words in the stream, we found that the word “curfew” appears in the stream at 2015-04-28 05:46:05, “@cbsbaltimore: A curfew is in place in \#Baltimore overnight from 10 p.m. to 5 a.m. this week.”, the word lockdown appears after an hour and 15 minutes in a different time interval of the word “curfew” after two widows, which means our system can capture relevant words that have not appeared in the
stream yet. The word “lockdown” appears at 2015-04-28 06:51:19, “Where the have the whole city on lockdown. No one let out and National guard or police on every block of the city”. The other word “nightfal” appears after 12 windows of the word “curfew” at 2015-04-28 08:59:38 , “Let's see what loot I get in this weeks nightfall! Post it if anything good!” and at 2015-04-28 16:40:08, “Hearing that \#Baltimore police made 235 arrests last night as nightfall approaches.” Finally, the word “impos” is the stemming of the word imposed appear in the stream after one window of the word “curfew” at 2015-04-28 06:02:13, “Baltimore schools are closed today and a 10 p.m. to 5 a.m. curfew will be imposed tonight.” All these examples prove that our system can enrich the static query with words that capture future events. It is worth noticing that the word "curfew" is also correlated with the word "violence" as appears in the table, this word ("violence") appears after 17 window from the word "curfew" at 2015-04-28 10:09:42, "You want to end the riots? Stop police violence! Stop bailing out man slaughtering officers! Stop police shooting people!." The top 3 nearest neighbor words for the word “looting” show the same trend as the top 3 words nearest neighbor of the word “curfew”.

%\textbf{AH: Here add a paragraph explaining what you observe in that table. What do we learn from these words? In what way are they correlated with the trigger words -- not technically, but substantively? Are these words present in the primary stream or haven't they appeared yet? If they haven't appeared yet, these would be great examples of the proactive nature of the query}

Algorithm 3 represents the process of the proposed method. The algorithm starts by adding the $d$ top-ranked DEC words  that do not appear in the top $k$ LDA words to the topic $t$ as explained in Algorithm 2. Then the function nearestNeighbors($V$,$w_{t}$,$i$) is used to find the $i$ nearest words closest to each word from topic $t$ in the vector space $V$. The resulting words are then saved in a list called $nearest$. For each topic $t$ we then attach the words that do not appear in topic $t$ from the $nearest$ list that is saved in $W_{v}$.  After adding the nearest neighbors words $W_{v}$ to topic $t$, we expand the query and return the results such that for a given document $s$ from  the primary stream $S_{1}$, if we can find any $l$ LDA, DEC, and nearest neighbor words in the document $s$ , then we add document $s$ into query result $q_{t}$. Finally, the aggregated  query results $Q$ for all topics will be returned by the end of the procedure.

\begin{table}[htbp]
\vspace{-.02in}
\caption{10-nearest neighbor words of word curfew and looting }\label{tab1}
\begin{center}
\begin{tabular}{|c|c|}
\hline
 Nearest Neighbor Words for "curfew"& Nearest Neighbor Words for "looting"\\
\hline
lockdown&vandal\\
\hline
nightfal&arson\\
\hline
impos&ransack\\
\hline
riot&quiktrip\\
\hline
loot&destruct \\
\hline
midnight&riot\\
\hline
polic&violenc\\
\hline
violence&protest\\
\hline
rioter&kill\\
\hline
protest&pillag\\
\hline

%\hline
\end{tabular}
\end{center}
\vspace{-.1in}
\end{table}

\begin{algorithm}
\caption{Query expansion using LDA words, DEC words, and Vector Space}

\begin{algorithmic}[1]
\linespread{1.30}\selectfont
\Variables
\State $Q$, Query result for all topics
\State $S_{1}$, Primary tweet stream
%\State $s$, A tweet in the primary tweet stream
\State $T$, A set of topics result from LDA
%\State $t$, A given topic in $T$
\State $w$, Time interval
\State $D$, DEC words for time interval $w$
\State $d$, Number of top-ranked DEC words to return
%\State $DEC_{w}$, Top-ranked DEC words that do not appear in the top LDA words
\State $V$, External semantic vector space
%\State $nearest$, A set of neighboring words of specific word
\State $i$, Number of nearest neighbor words to return
%\State $q_{t}$, Query result associated with a given topic $t$
\EndVariables

\Procedure{Proactive Vector Space}{
$Q$,$T$,$V$,$w$,$S_{1}$,$d$,$D$,$i$}   
    \State $Q$ $\leftarrow$ $\emptyset$ 
    \State $W_{v}$ $\leftarrow$ \big[ \big]
     \For{$t$ in $T$} \Comment{for each topic $t$ return the query result associated with it } 
        \State $DEC_{w} \leftarrow $$DEC$($w$,$t$,$d$, $D$) \Comment{return $d$ top-ranked DEC words that do not appear in the top LDA words of topic $t$ for time interval $w$ }
         \State $t$ $\leftarrow$	$t$ $\cup$ 	$DEC_{w}$ \Comment{attach top $d$ DEC words to the topic $t$}
        \State $qt$ $\leftarrow$ $\emptyset$ \Comment{initialize query result associated with topic $t$}
        \For{$w_{t}$ in $t$}  \Comment{for each word $w_{t}$ in $t$ return $i$ nearest neighbor words }
             \State $nearest$ $\leftarrow$
            nearestNeighbors($V$,$w_{t}$,$i$) \Comment{return $i$ nearest neighbor words to the word $w_{t}$ that do not appear in the LDA and DEC words of topic $t$ for time interval $w$}
            \For{$newWord$ in $nearest$} 
             \If{$newWord$ not in $t$ } \Comment{for each nearest neighbor word check if it are already in $t$ }
                    \State $W_{v}$ $\leftarrow$	$W_{v}$ $\cup$ 	$newWord$  
             \EndIf 
             \EndFor  
         \EndFor
         \State $t$ $\leftarrow$	$t$ $\cup$ 	$W_{v}$ \Comment{attach nearest neighbor words to $t$}
          \For{$s$ in $S_{1}$} \Comment{for each tweet $s$ in $S_{1}$ check if the tweet has the LDA, DEC, and nearest neighbor words}
             \State $cnt$=0 \Comment {count the number of LDA,DEC, and nearest neighbor words in the tweets $s$ }
             \For{$v$ in $t$} \Comment {for each word in the topic $t$ check if the word $v$ in the tweet $s$}
                \If{$v$ in $s$}
                    \State $cnt$+=1 \Comment{increase the number of of LDA  DEC and nearest neighbor words in tweets $s$ }
                \If{$cnt$ $\geq$ $l$} \Comment {if $cnt$ $\geq$ the length of the query $l$, add the tweet $s$ to the $q_{t}$ } 
                    \State $q_{t}$ $\leftarrow$	$q_{t}$ $\cup$ 	$s$ \Comment{ add the tweet $s$ to the query result $q_{t}$  }
                \EndIf
                \EndIf
            \EndFor  
    \EndFor 
    \State $Q$ $\leftarrow$	$Q$ $\cup$ 	$q_{t}$
    \EndFor
    \State return $Q$
\EndProcedure
\end{algorithmic}
\end{algorithm}

\textbf{Proactive Co-occurrence} Query expansion using LDA words, DEC words, and Co-occurrence frequency. This method is the same as the previous method except that the most relevant words in the external data is defined using the words’ highest frequency. This method returns a set of words that have the highest number of occurrences for a certain LDA and DEC word.  We built a dictionary $F$ that consist of bi-grams (pair of adjacent words) from the external source, then we compute the frequency for all the bi-grams in the external data to return the $j$ highest word frequency related to each LDA and DEC word. Table \ref{tab2} presents the top 10 highest number of co-occurrence words for the two words "curfew" and "looting". As appear in the table, the top 3 highest number of co-occurrence words for the word “curfew” are: “militari”, “ impos”, and “nationwid”. As we mentioned before the word “curfew” appears in the stream at 2015-04-28 05:46:05. The word “militari” appears after 20 minutes in a different time interval of the word “curfew” after one window. which proves that our system can capture relevant words that have not appeared in the stream yet using the words co-occurrence. The word “militari” appears at 2015-04-28 06:06:53, “Don't forget. The heavily militarized police presence formed in response to PEACEFUL student protests. Not  a violent one.” The other word “impos” is the same word that returned using the proactive vector space which means this word has the highest words co-occurrence and the highest cosine similarity to the word “curfew”. Finally, the word “nationwid” appears in the stream after one window of the word “curfew” at 2015-04-28 06:00:13, "The police are ridiculous to EVERYBODY. Most people can agree with that. The police force needs to change nationwide." The top 3 highest number of occurrences words for the word “looting” show the same trend as the highest words co-occurrence for the word “curfew”.

%\textbf{AH: Same as above. What do we learn from this table?}

Algorithm 4 represents the process of the proposed method. The algorithm starts by adding the $d$ top-ranked DEC words  that do not appear in the top $k$ LDA words to the topic $t$ as explained in Algorithm 2. The Algorithm then identifies the $j$ highest frequency words for each word in topic $t$. The Function highestFreq($F$,$w_{t}$,$j$) is then used to attach the highest frequency words $W_{f}$ to each topic $t$, with the resulting words saved in a list called $freq$. For each topic $t$, we then attach the words  that do not appear in topic $t$ from $freq$ list that is saved in $W_{v}$. After adding the highest frequency words $W_{v}$ to topic $t$, we expand the query  and return the results such that for a given document $s$ from  the primary stream $S_{1}$, if we can find any $l$ LDA, DEC, and highest frequency words in the document $s$, then we add document $s$ into query result $q_{t}$.Finally, the aggregated  query results $Q$ for all topics are returned by the end of the procedure. 

%\textbf{AH: Again, what does "combination" here mean?}

\begin{table}[htbp]
\vspace{-.02in}
\caption{10 Highest number of occurrences words of the two words "curfew" and "looting" }\label{tab2}
\begin{center}
\begin{tabular}{|c|c|}
\hline
 Highest Word Frequency  for "curfew"&   Highest Word Frequency for  "looting"\\
\hline
militari&secur\\
\hline
impos&destruct\\
\hline
nationwid&extens\\
\hline
overnight&sporad\\
\hline
hour&systemat \\
\hline
violat&store\\
\hline
began&riot\\
\hline
mandatori&violenc\\
\hline
citywid&widespread\\
\hline
citi&vandal\\
\hline

%\hline
\end{tabular}
\end{center}
\vspace{-.1in}
\end{table}
\begin{algorithm}
\caption{Query expansion using LDA words, DEC words, and Co-occurrence frequency}

\begin{algorithmic}[1]
\linespread{1.30}\selectfont
\Variables
\State $Q$, Query result for all topics
\State $S_{1}$, Primary tweet stream
%\State $s$, A tweet in the primary tweet stream
\State $T$, A set of topics result from LDA
%\State $t$, A given topic in $T$
\State $w$, Time interval
\State $D$, DEC words for time interval $w$
\State $d$, Number of top-ranked DEC words to return
%\State $DEC_{w}$, Top-ranked DEC words that do not appear in the top LDA words
\State $F$, External bi-grams dictionary
%\State $freq$, A set of highest frequency words of specific word 
\State $j$, Number of nearest neighbor words to return for a specific word
%\State $q_{t}$, Query result associated with a given topic $t$
\EndVariables
\Procedure{Proactive Co-occurrence}{$Q$,$T$,$F$,$w$,$S_{1}$,$d$,$D$,$j$}   
    \State $Q$ $\leftarrow$ $\emptyset$ 
    \State $W_{f}$ $\leftarrow$ \big[ \big]
     \For{$t$ in $T$}  \Comment{for each topic $t$ return the query result associated with it }
        \State $DEC_{w} \leftarrow $$DEC$($w$,$t$,$d$, $D$) \Comment{return $d$ top-ranked DEC words that do not appear in the top LDA words of topic $t$ for time interval $w$ } 
         \State $t$ $\leftarrow$	$t$ $\cup$ 	$DEC_{w}$ \Comment{attach top $d$ DEC words to the topic $t$}
        \State $qt$ $\leftarrow$ $\emptyset$ \Comment{initialize query result associated with topic $t$}
        \For{$w_{t}$ in $t$}  \Comment{for each word $w_{t}$ in $t$ return $j$ highest frequency words }
             \State freq $\leftarrow$
            highestFreq($F$,$w_{t}$,$j$)  \Comment{return $j$ highest frequency words to the word $w_{t}$ that do not appear in the LDA and DEC words of topic $t$ for time interval $w$}
            \For{$newWord$ in freq}
             \If{$newWord$ not in $t$ } \Comment{for each highest frequency word check if it are already in $t$ }
                    \State $W_{v}$ $\leftarrow$	$W_{v}$ $\cup$ 	$newWord$ 
             \EndIf 
             \EndFor  
         \EndFor
         \State $t$ $\leftarrow$	$t$ $\cup$ 	$W_{v}$ \Comment{attach highest frequency words to $t$}
          \For{$s$ in $S_{1}$} \Comment{for each tweet $s$ in $S_{1}$ check if the tweet has the LDA,  DEC, and highest frequency words}
             \State $cnt$=0 \Comment {count the number of LDA, DEC, and highest frequency words in the tweets $s$ }
             \For{$v$ in $t$} \Comment {for each word in the topic $t$ check if the word $v$ in the tweet $s$}
                \If{$v$ in $s$}
                    \State $cnt$+=1\Comment{increase the number of of LDA  DEC and nearest neighbor words in tweets $s$ }
                \If{$cnt$ $\geq$ $l$}\Comment {if $cnt$ $\geq$ the length of the query $l$, add the tweet $s$ to the $q_{t}$ } 
                    \State $q_{t}$ $\leftarrow$	$q_{t}$ $\cup$ 	$s$\Comment{ add the tweet $s$ to the query result $q_{t}$  }
                \EndIf
                \EndIf 
            \EndFor  
    \EndFor 
    \State $Q$ $\leftarrow$	$Q$ $\cup$ 	$q_{t}$
    \EndFor
    \State return $Q$
\EndProcedure
\end{algorithmic}
\end{algorithm}

\section{Evaluation}
 In this section, we will give an overview of the evaluation of our query expansion methods. First, we describe the data that we use to detect the emergent topics and expand the query. Second, we give a detailed description of the evaluation of our query expansion methods. 
 
\subsection{Data Description}
We use Twitter data collected for one specific public safety event: the 2015 Baltimore protests in response to the death of a Baltimore resident Freddie Gray. The death of Gray caused a series of protests and violence, which led to a whole city curfew on the evening of April 28th. 

We purchased archived tweets from Gnip, a company that provides access to the full archive of public Twitter data. We used broad search words to collect tweets in order to create a noisy data stream that covers 
tweets related to the Baltimore events as well as unrelated events.\footnote{Our data set was collected with the following search words: joseph kent, freddie gray, eric garner, ferguson, curfew, police, riot, protests, loot, looting, \#purge, \#baltimore, \#baltimoreriots, \#baltimoreuprising, \#freddiegray, \#josephkent, \#blacklivesmatter, \#onebaltimore, rioter, charge, charged, murder, homicide, mosby, corporal, \#mayday, justice, \#blackspring, \#freddiegray’s, cops, unjustified, spinal, broken spine, arrested, thugs, thug, \#marilynmosby, \#wakeupamerica, freddie, racist, racism, \#baltimoreprotest, propaganda, officers, knife.  A logical OR expression was used to filter the words, i.e., for example, by keeping the term Baltimore, we obtained all tweets related to the city, and not necessary to the event \cite{avudaiappan2017detecting}. The Tweet data is preprocessed before applying any query expansion methods; we removed the stop words, URLs, numbers, and all non-English tweets.} 

Our data set comprises 20.5 million tweets covering fifteen days from April 17th to May 3rd, 2015. Because of its noisy nature, the stream is ideal to evaluate our method’s ability to detect emergent topics and expand the query. 

For our external source, we decided to use an archival news article published one year before the Baltimore events in order to predict the occurrence of relevant words that have not appeared in the stream yet. Using an archival external source allows our system to search for old words that related to a static query to augment it with correlated information to predict future events or topics. We chose New York Times (NYT) and CNN as our source of external data because these sources have a public API that can be used to crawl the archived news articles. We obtained 30,456 articles from NYT and 14,145 from CNN.

%\textbf{AH: The last paragraph needs MUCH MORE explanation. Why one year before the events? Why CNN and NYT? How many articles were included? How many words? etc.}

\subsection{Evaluation}
How good is our proposed  query expansion method in expanding queries and predicting future topics? In this section, we answer this question by conducting two types of experiments to evaluate our methods using Twitter data from the 2015 Baltimore protests. For both experiments, we simulate real-time stream processing by constructing a primary stream from the full data. This primary stream consists of all tweets that contain the word “police”, a total of 5.1 million tweets. We divide this primary stream into $15$-minute time intervals, which we call “windows”. On each window, we use the DEC metric to determine during which windows an emergent event occurred. More precisely,  we calculate the Jaccard similarity ($J_{s}$) between the top $200$ DEC words between the current window and the 3-previous windows. If this similarity is less than or equal to $15$\%, we assume that an emergent event has occurred. We chose low Jaccard similarity because it indicates we have found relatively unique set of emergent words and these words have not appeared in the stream yet. Using this metric, the primary stream resulted in 373 windows with emergent events out of the 1573 windows. Then LDA was used to extract a set of 5 topics in the targeted time window (intervals have emergent events). Then, we use the top 20 words from each topic to form the top-ranked words. 

In order to relate our results to reality, we identified three key events from timelines published by news outlets \cite{knuthwebsite} to pick some time intervals to use in our experiments. Therefore, in addition to the first-time interval triggered by our algorithm (time interval 16), we used the time intervals 155, 781, and 1065 based on the events that happened in Baltimore. Time interval 155 at 7:00 am, April 19, captures the tweets about the death of Freddie Gray. Time interval 781 on April 25, has the tweets about looting, violence, and protest. Time interval 1065 at 10:00 pm on April 28, has the Baltimore curfew tweets. Each experiment is applied at these time intervals.
The two experiments are explained in the following subsections:

%\subsection{Experiment 1 } 
\subsection{Experiment 1: Quantity and Quality of Retrieved Data}
Experiment 1 compares the performance of our proactive query expansion methods  proactive VS and proactive CO with respect to the reference methods static and emergent using streaming data quality indicators for a certain time interval. Each method returns a set of tweets called query result (Q).  The following quality indicators metrics are: 

{\it Volume (measured by tweet count):} This metric finds the total number of tweets matching a specific query condition from a certain time interval to the end of the primary stream.

{\it Relevance (measured by hashtag count):} This metric finds the total number of hashtags in the tweets matching a specific query condition from a certain time interval to the end of the primary stream.

\begin{figure*}[ht]
   \includegraphics[width=\textwidth]{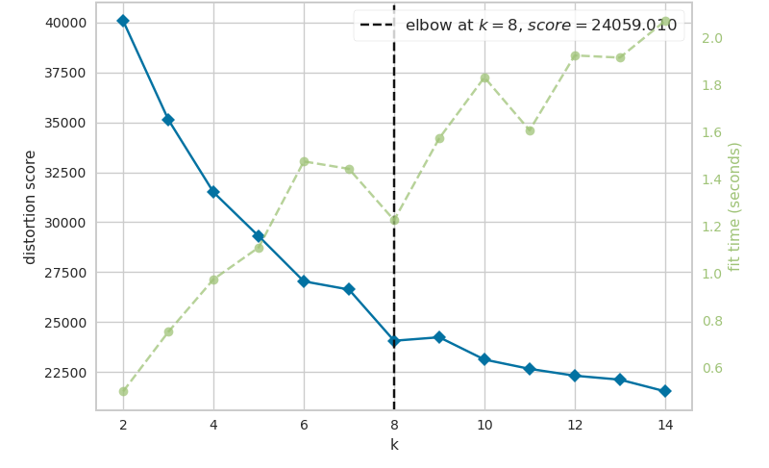}
   \caption{Elbow Method}
   \label{elbw}
   \end{figure*}

{\it Conciseness (measured through hashtags clustering):} This metric clusters the tweets matching a specific query condition from a certain time interval to the end of the stream. In order to determine how concise the stream is, we use the hashtags to check if the query results return similar hashtags based on the number of clusters needs to cluster them. We find the number of clusters using k-means  \cite{macqueen1967some}. K-means is used to cluster the query results for each method such that: the data points are the set of tweets, and the features of the cluster are the hashtags. The lower the optimal number of clusters, the more concise and specific the stream is.  To find the optimal number of clusters (k), we used the elbow method \cite{bengfort_yellowbrick_2018,susan2020learning}. Where the optimal number of clusters is represented in a graph as is an inflection point (elbow) using the average distortion score. The distortion score is the sum of squared differences of each point to its assigned center. In this experiment, the distortion score is computed from k = 2 to k = 15 clusters.  Fig.~\ref{elbw} shows the graphical representation of the elbow test, the inflection point or the elbow in this test is 8 which is the optimal number of clusters.

The higher the tweets count and hashtag count, the better and more relevant result is and the lower number of clusters, the more concise and specific the result is.

%\subsection{Experiment 2 }
\subsection{Experiment 2: Predictive Power of Retrieved Data}
Experiment 2 tests the effectiveness of our proposed methods to predict future emerging events or to predict the conversations from previous time intervals using a different set of hashtags.  For example, we know that in time interval $x+n$ there was a curfew. Are our methods able to predict the curfew from time interval $x$ which is related to protests? Will the protest lead to curfew? We picked and used a different set of hashtags matching the events that happened in Baltimore to determine if our methods can predict them from previous time intervals. The question here is which hashtag to use and how to validate? We used two categories of hashtags, the highest frequency hashtag, and the lowest frequency hashtags. In this experiment, we used a different set of hashtags that are related to some events that happened in Baltimore in response to the death of Freddie Gray. And for the validation, we used precision as an indicator of the effectiveness of our methods in predicting events. The precision is defined as the count of a certain hashtag from the query results divided by the total number of the same hashtag from the start of a certain time interval  to the end of the stream.

\section{Results }
This section compare the results of the four methods based on volume, relevance, and conciseness which we defined above.

\subsection{Experiment 1 }
In this experiment, we compared the performance of  proactive VS and proactive CO with respect to the reference methods static and emergent using streaming data quality indicators; tweet count, hashtag count, and hashtag clustering, for  different time intervals (16, 155, 781, and 1065). 

\begin{figure*}[ht]
   \includegraphics[width=\textwidth]{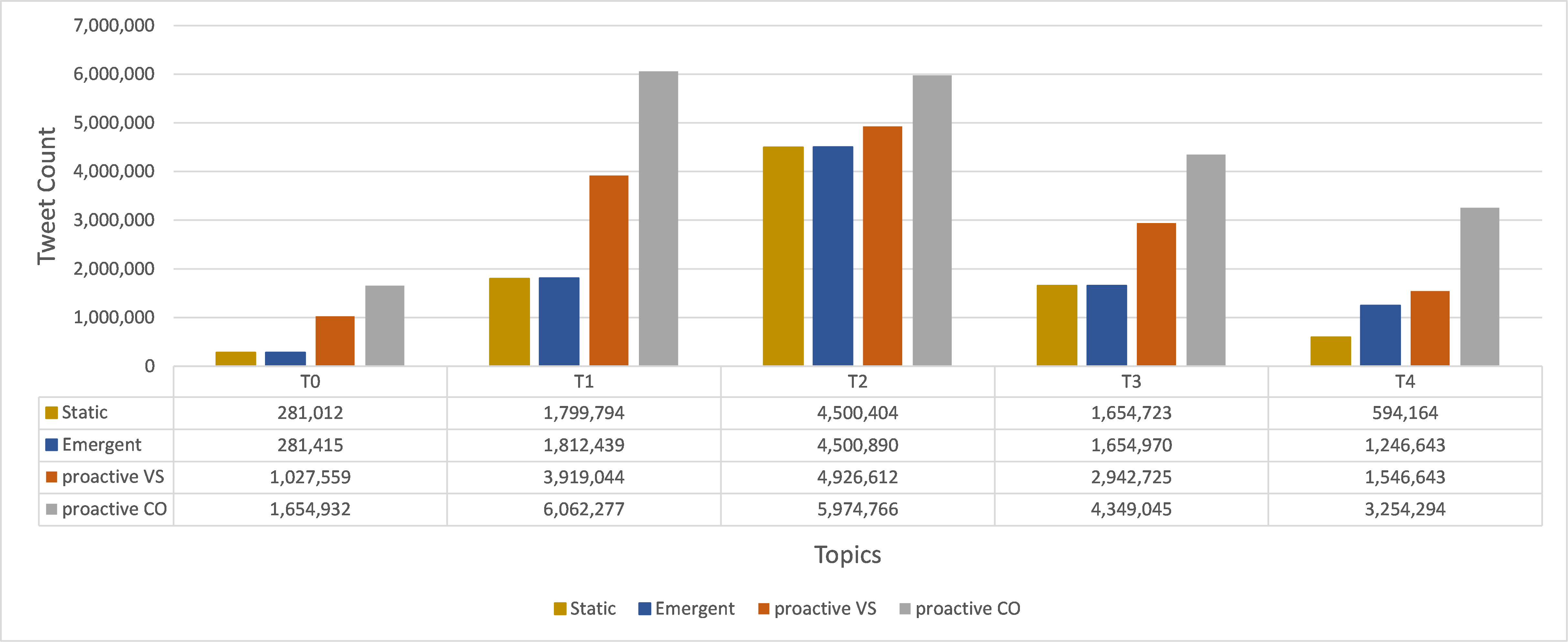}
   \centering
   \caption{Tweet Count Over Time interval 16 of Query Results}
   \label{tc16}
   \end{figure*}

\begin{figure*}[ht]
   \includegraphics[width=\textwidth]{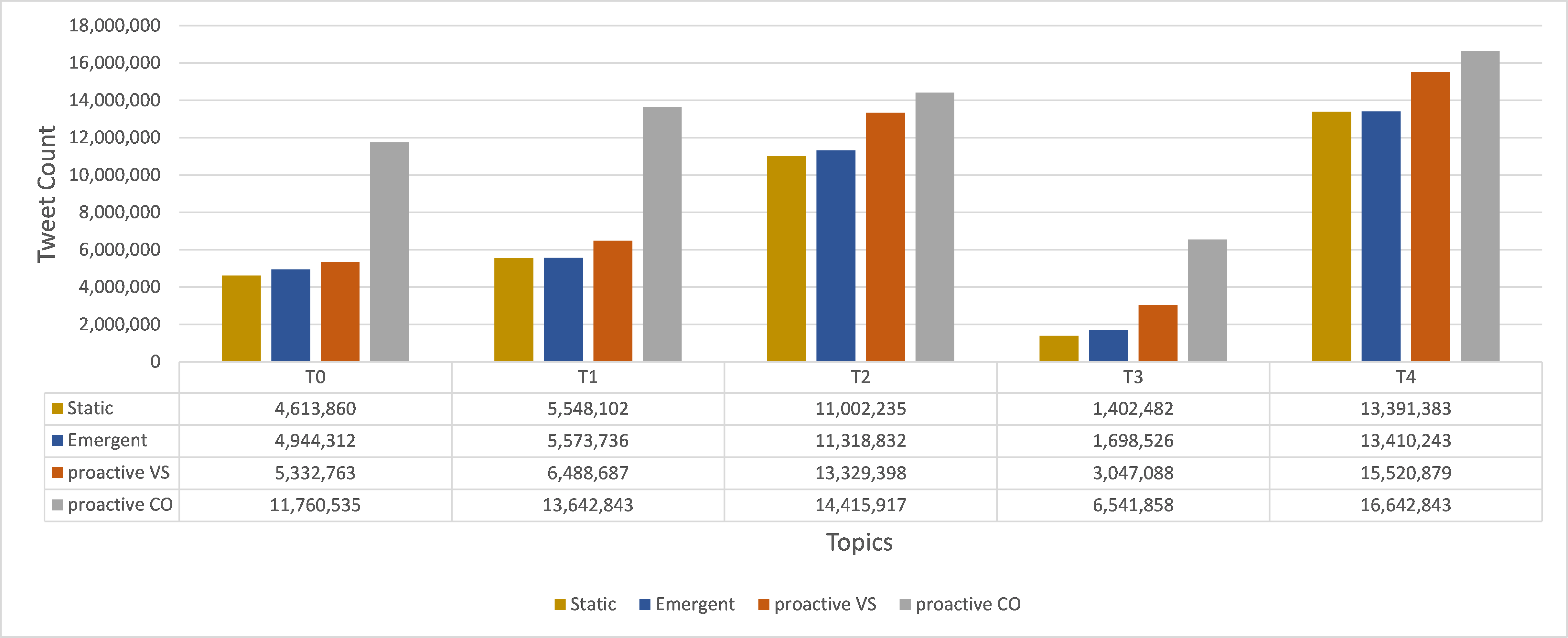}
   \centering
   \caption{Tweet Count Over Time interval 155 of Query Results }
   \label{tc155}
   \end{figure*}

\begin{figure*}[ht]
   \includegraphics[width=\textwidth]{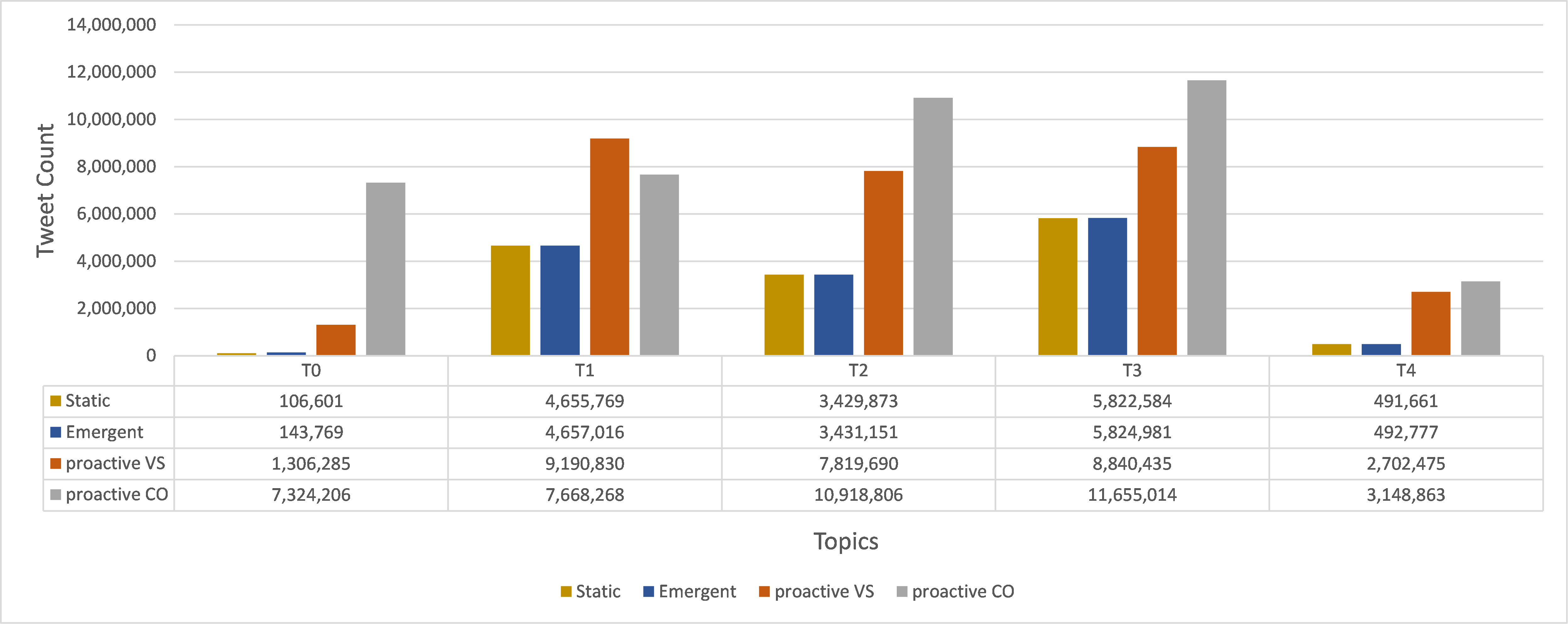}
   \centering
   \caption{Tweet Count Over Time interval 781 of Query Results }
   \label{tc781}
   \end{figure*}

\begin{figure*}[ht]
   \includegraphics[width=\textwidth]{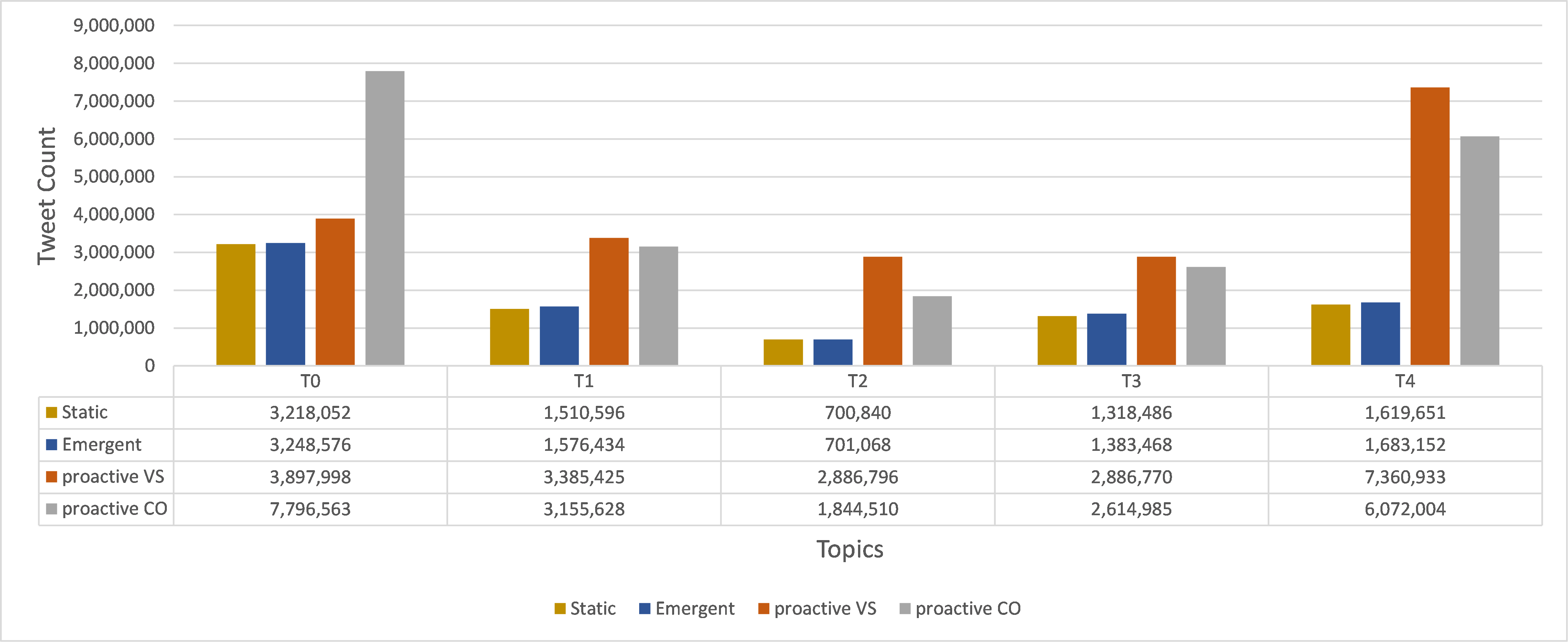}
   \centering
   \caption{Tweet Count Over Time interval 1065 of Query Results }
   \label{tc1065}
   \end{figure*}

In terms of volume (i.e, tweet count), figures \ref{tc16}, \ref{tc155}, \ref{tc781}, and \ref{tc1065} show the number of tweets for the query results per five topics using the four methods at time intervals 16, 155, 781, and 1065, respectively. For all time intervals, we found that proactive VS and proactive CO  significantly outperform emergent and static methods. In other words, proactive VS and proactive CO can return more tweets than the others per each topic, which means adding new words from external sources covers more data.  For instance, at time interval 16 under the topic (T0), proactive VS and proactive CO returned 3.65 and 5.88 times more tweets than static, respectively. Also, proactive CO returns more tweets than any other method which proves its efficiency. 

In Figure. \ref{tc155}, it can be seen that topics two and four covers more tweets than other topics per all methods at time interval 155. It indicates that there is an important event that happened at this time interval. To further investigate why these two topics have the highest number of tweets, we looked at the LDA words for these topics (T2, and T4). As a result, the LDA words included: Freddie Gray, died, killed, black, arrested, beaten, and officer. These words were the majority of the tweets in our stream at that time.

\begin{figure*}[ht]
   \includegraphics[width=\textwidth]{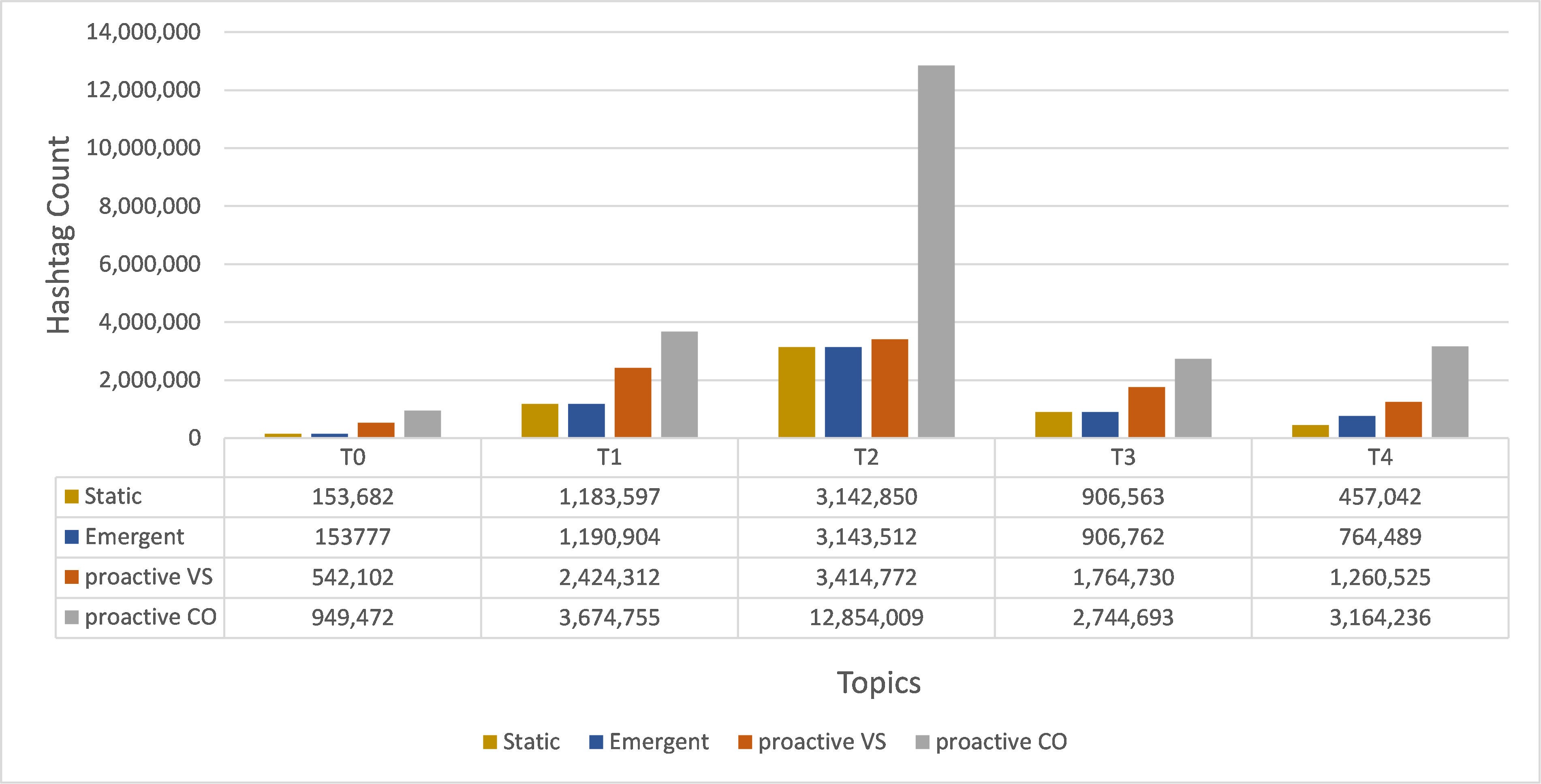}
   \centering
   \caption{Hashtag Count Over Time interval 16 of Query Results }
   \label{h16}
   \end{figure*}

\begin{figure*}[ht]
   \includegraphics[width=\textwidth]{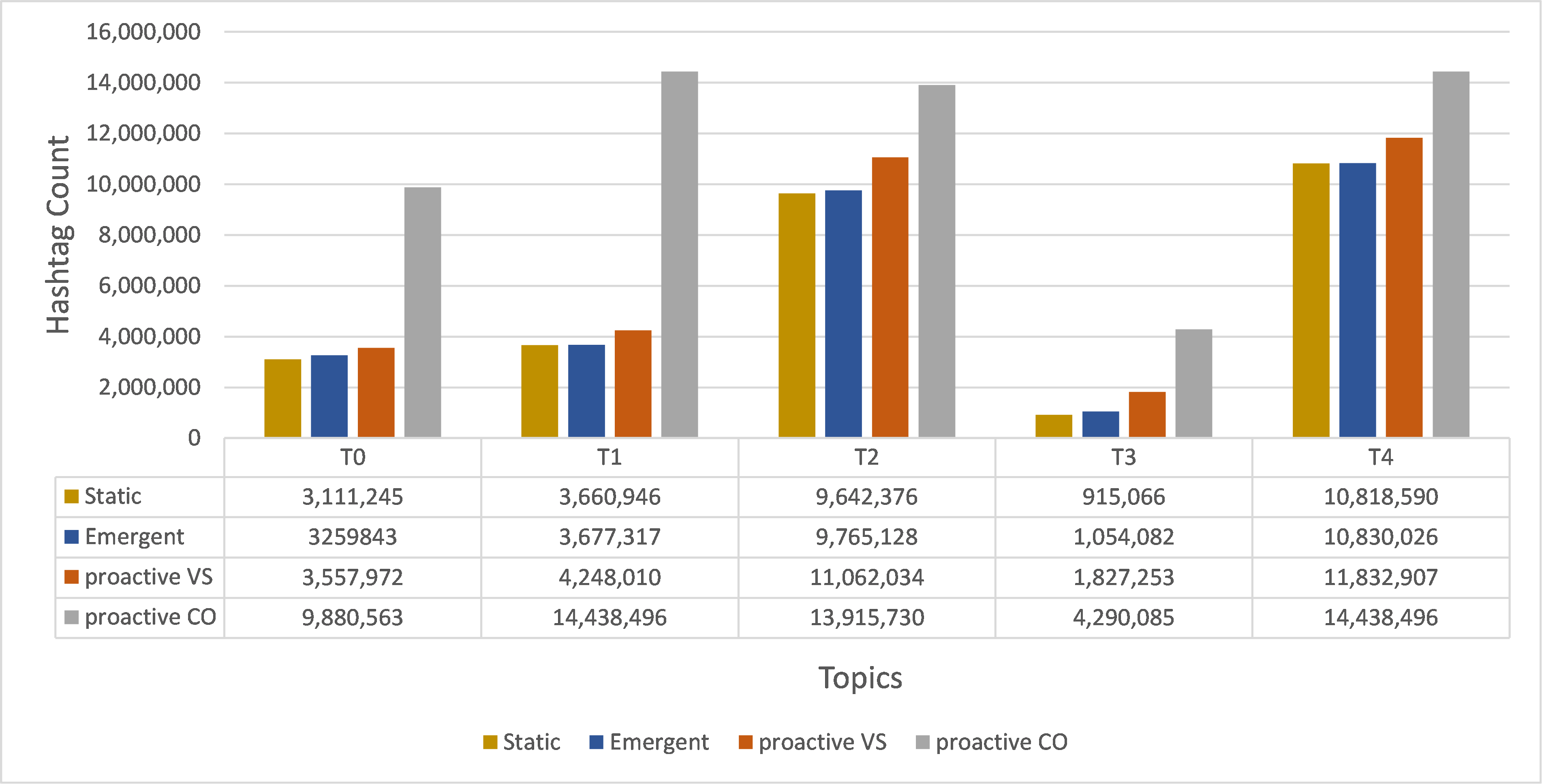}
   \centering
   \caption{Hashtag Count Over Time interval 155 of Query Results }
   \label{h155}
   \end{figure*}

\begin{figure*}[ht]
   \includegraphics[width=\textwidth]{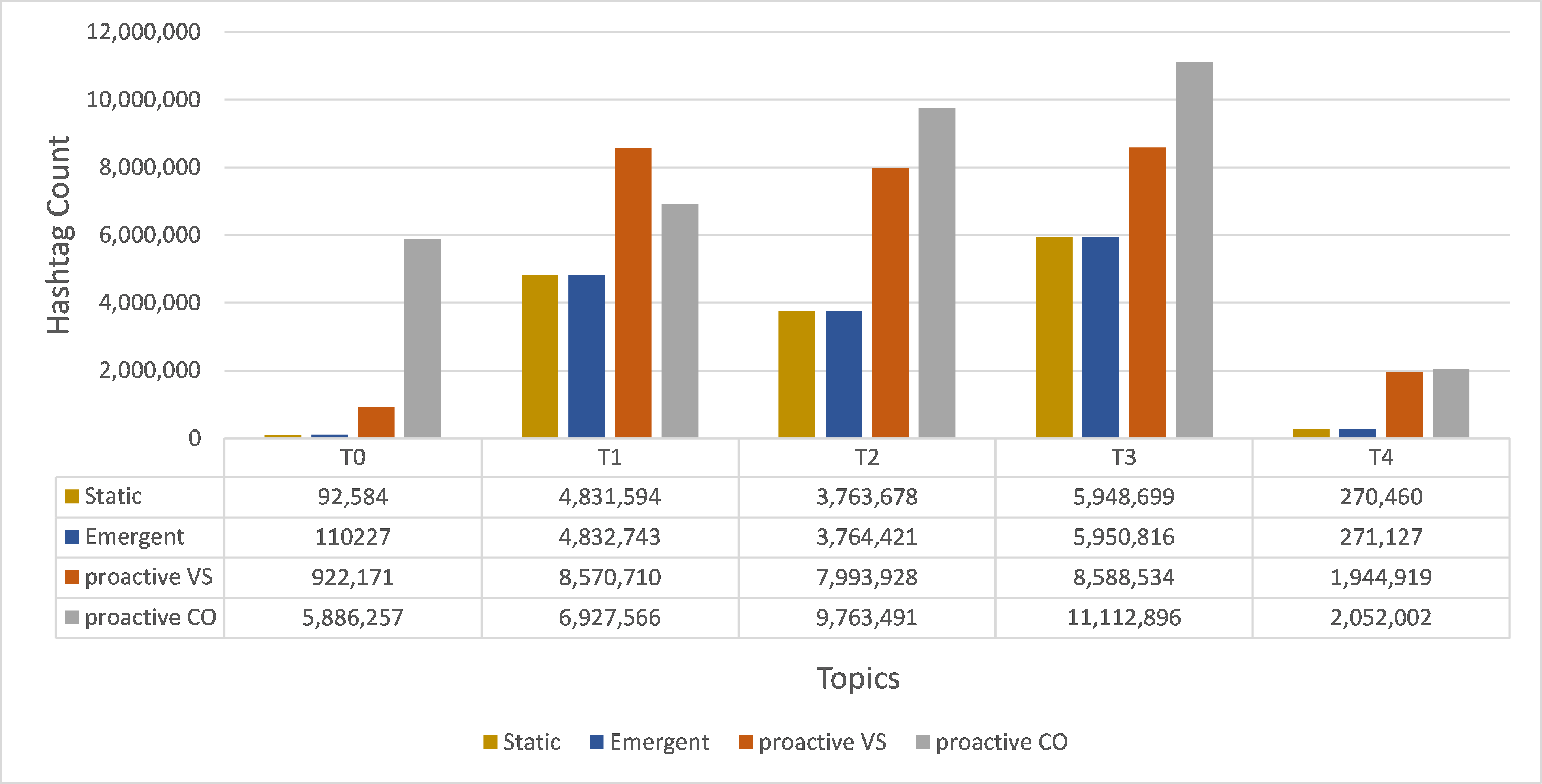}
   \centering
   \caption{Hashtag Count Over Time interval 781 of Query Results }
   \label{h781}
   \end{figure*}

\begin{figure*}[ht]
   \includegraphics[width=\textwidth]{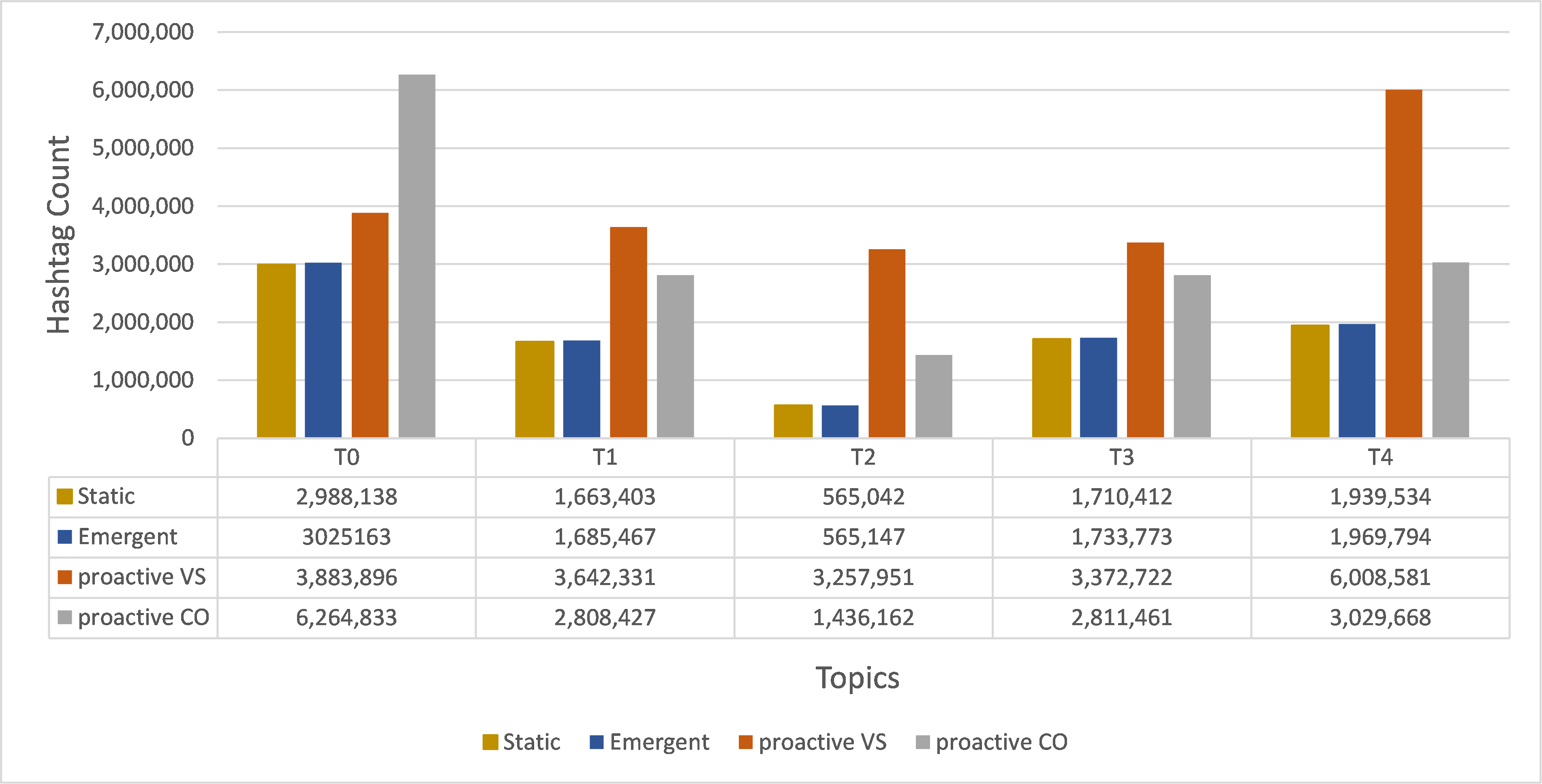}
   \centering
   \caption{Hashtag Count Over Time interval 1065 of Query Results }
   \label{h1065}
   \end{figure*}

In terms of relevance (i.e, hashtag count),  figures  \ref{h16}, \ref{h155}, \ref{h781}, and \ref{h1065} show the number of hashtags for the query result associated with each of the five topics  and generated using the four methods for all time intervals. As we expected, a higher number of hashtags is associated with proactive VS and proactive CO comparing with static and emergent methods. 

\begin{figure*}[ht]
   \includegraphics[width=\textwidth]{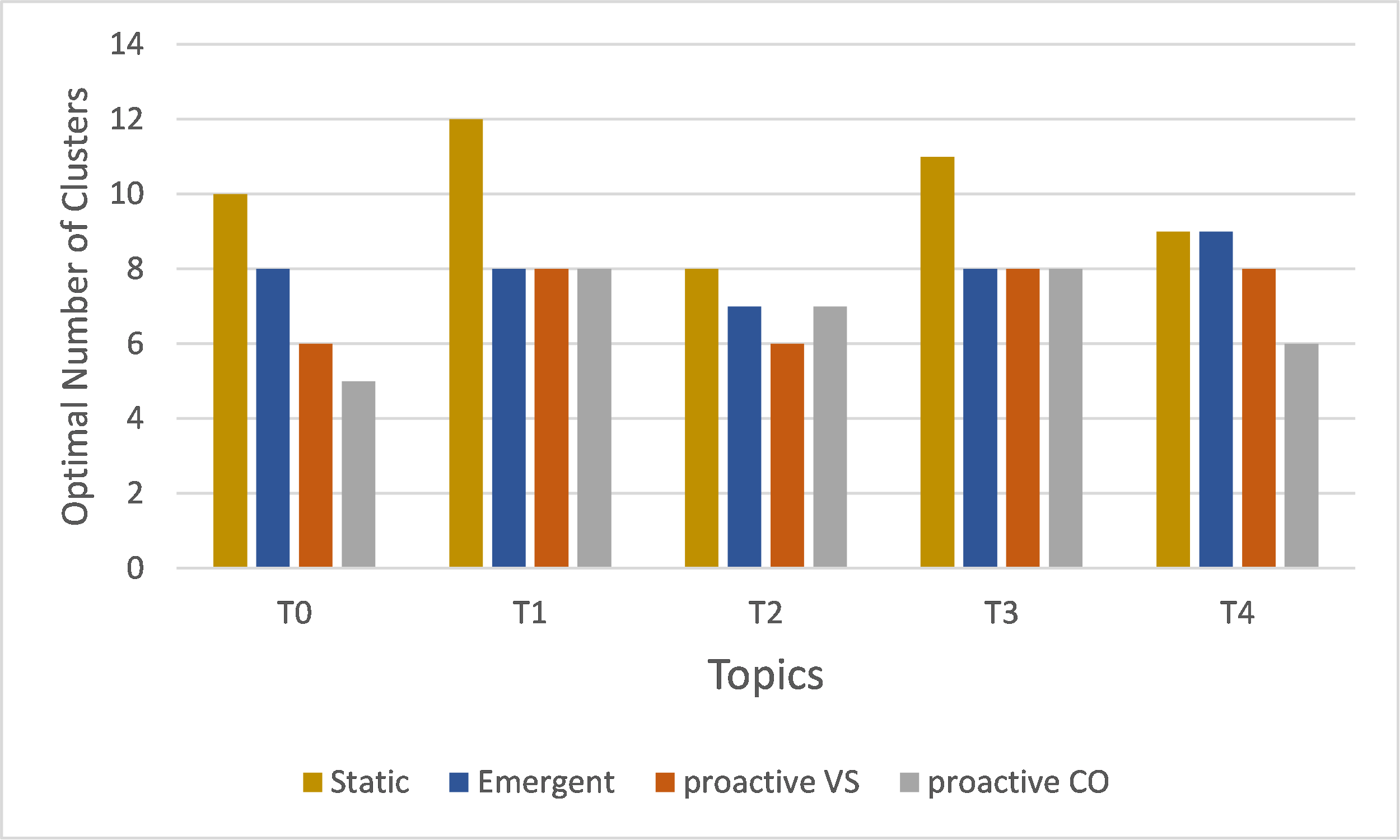}
   \centering
   \caption{Optimal Number of Clusters Over Time interval 16 of Query Results }
   \label{opt16}
   \end{figure*}

\begin{figure*}[ht]
   \includegraphics[width=\textwidth]{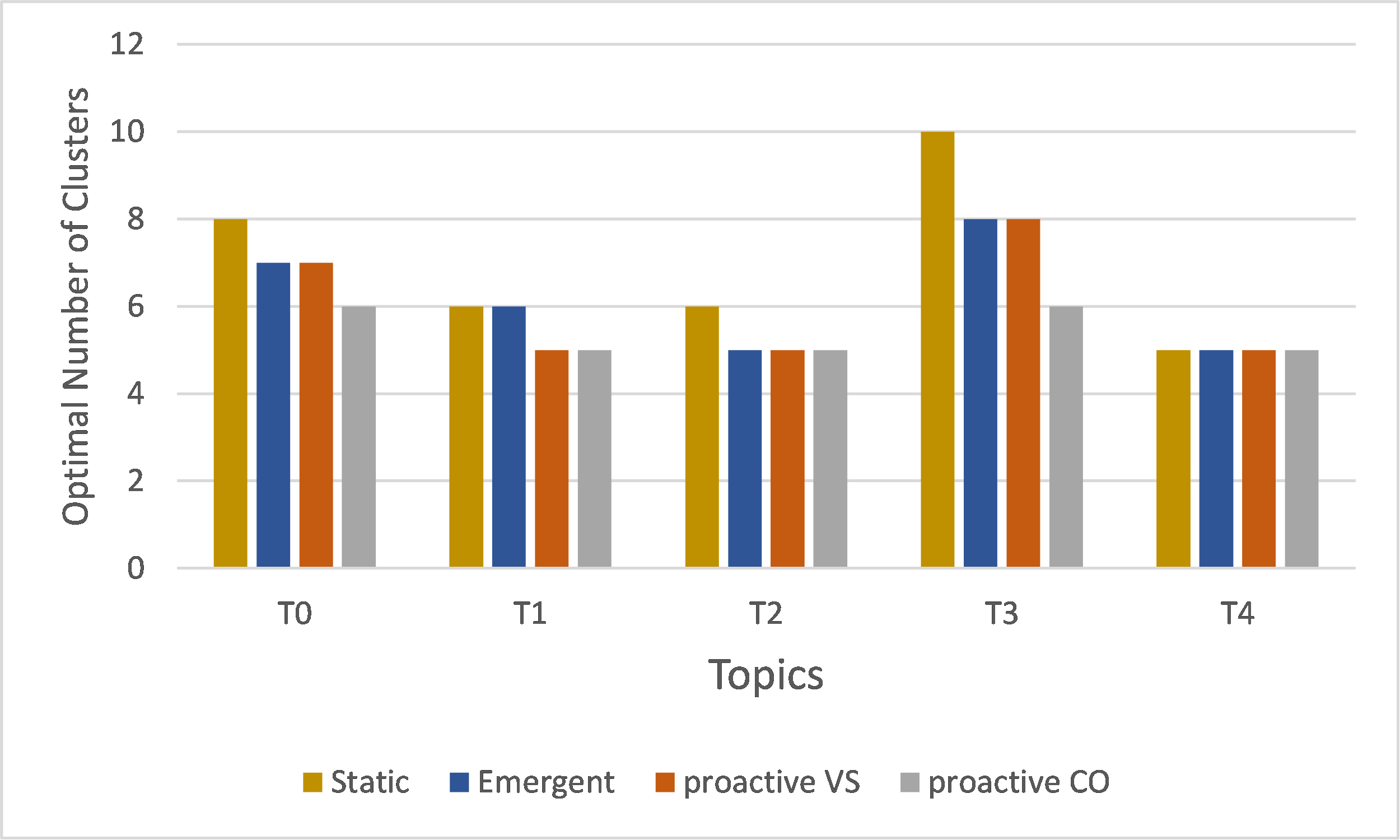}
   \centering
   \caption{Optimal Number of Clusters Over Time interval 155 of Query Results }
   \label{opt155}
   \end{figure*}

\begin{figure*}[ht]
   \includegraphics[width=\textwidth]{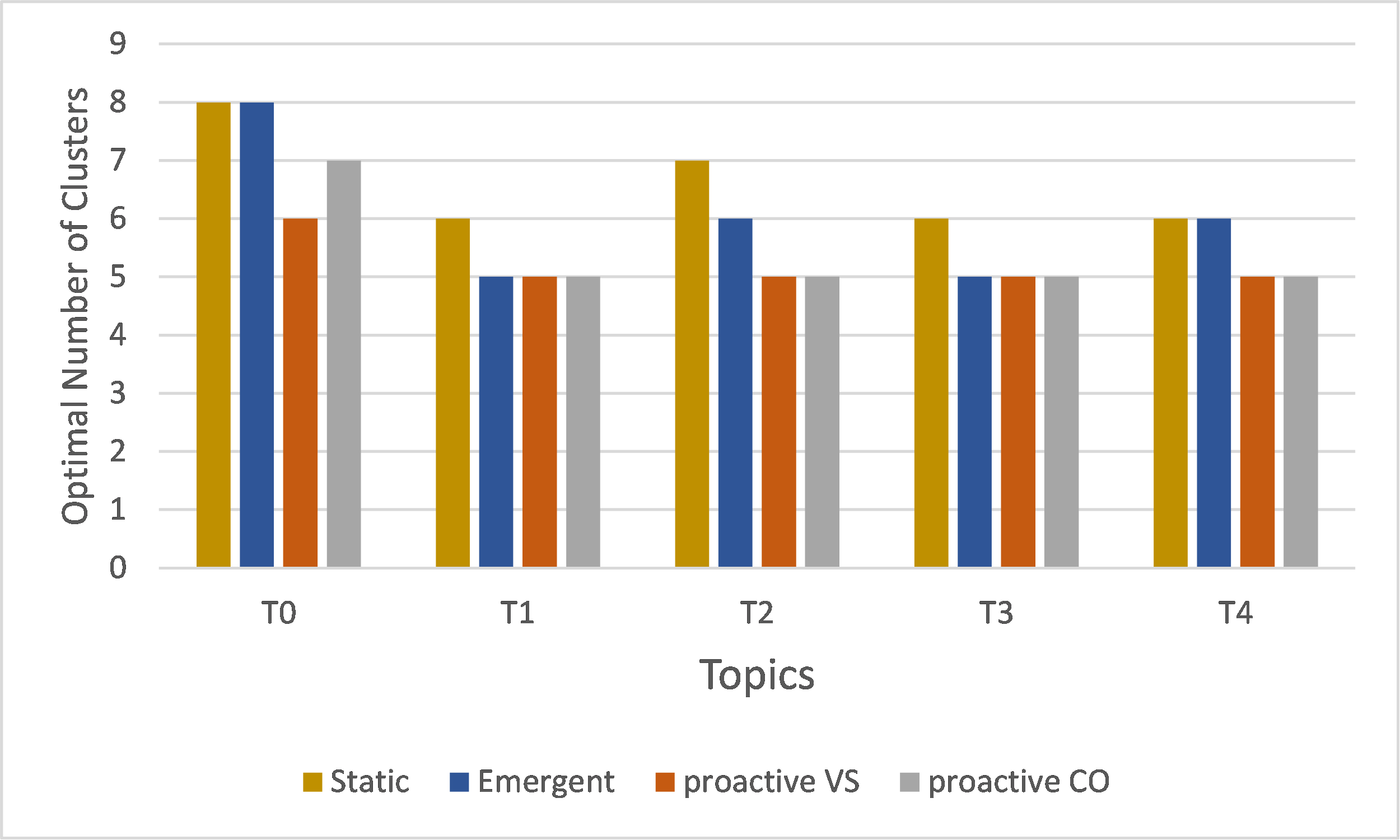}
   \centering
   \caption{Optimal Number of Clusters Over Time interval 781 of Query Results }
   \label{opt781}
   \end{figure*}

\begin{figure*}[ht]
   \includegraphics[width=\textwidth]{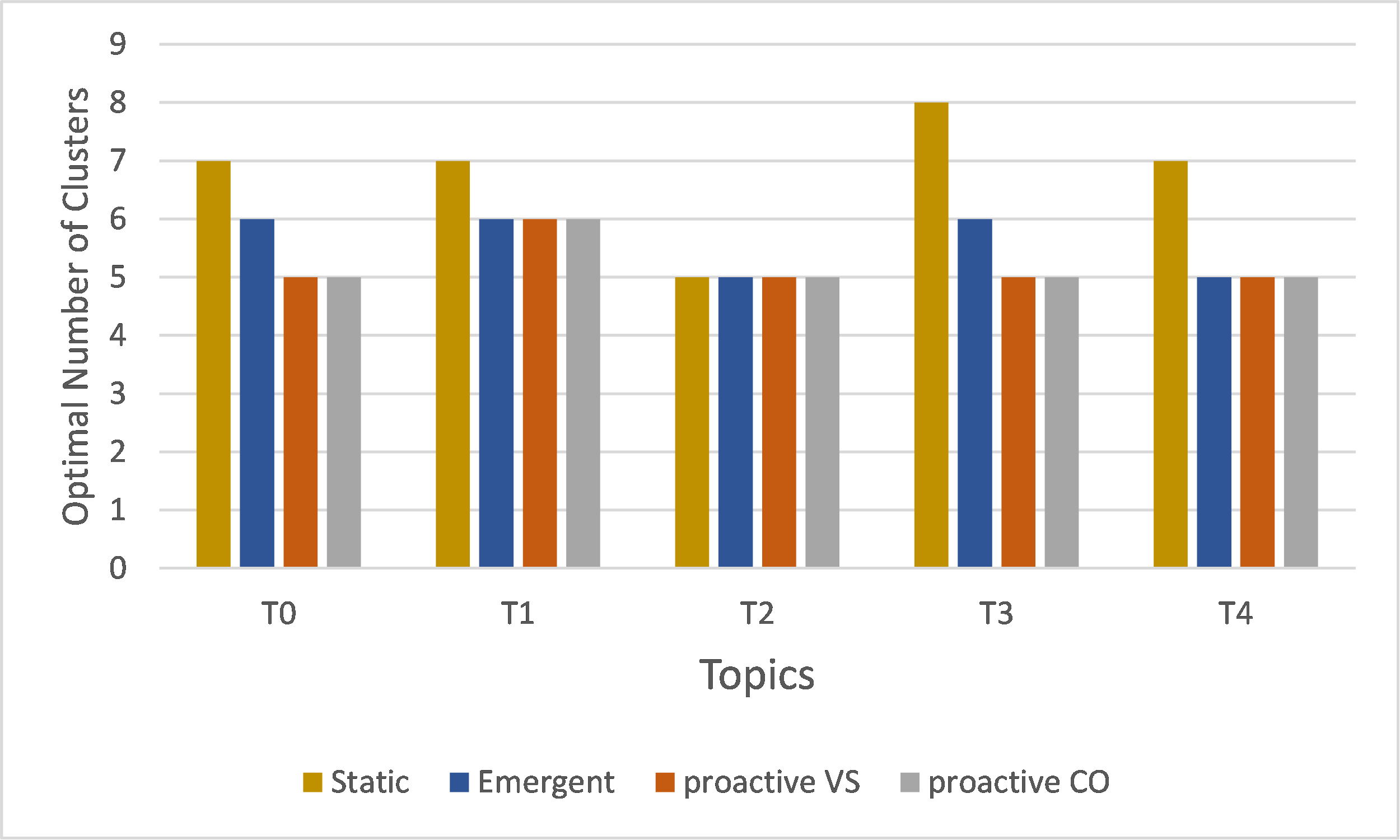}
   \centering
   \caption{Optimal Number of Clusters Over Time interval 1065 of Query Results }
   \label{opt1065}
   \end{figure*}

In term of conciseness (i.e, quality of hashtag clustering), figures\ref{opt16}, \ref{opt155}, \ref{opt781}, \ref{opt1065} show the optimal number of clusters $k$ for the query result using the k-means  elbow method per each of the five topics per each of the four methods for all time intervals. For all topics and all-time intervals, proactive VS and proactive CO return more concise tweets, despite the fact they return more tweets than other methods.

For example, In Figure.\ref{opt16} for topic 0 at time interval 16, we can cluster the tweets using 6 and 5 clusters using proactive VS and proactive CO respectively, while we need 10 and 8 clusters to cluster them using  static and emergent respectively.

%\begin{figure}
 %    \centering
  %   \begin{subfigure}[b]{0.5\textwidth}
   %      \centering
    %     \includegraphics[width=\textwidth]{16_optk_new2_new.png}
     %    \caption{Time Interval 16 }
      %   \label{TimeInterval16}
     %\end{subfigure}
     %\hfill
     %\begin{subfigure}[b]{0.5\textwidth}
      %   \centering
       %  \includegraphics[width=\textwidth]{155_optk_new2_new.png}
        % \caption{Time Interval 155}
        % \label{TimeInterval155}
     %\end{subfigure}
     %\hfill
     %\begin{subfigure}[b]{0.5\textwidth}
      %   \centering
       %  \includegraphics[width=\textwidth]{781_optk_new2_new.png}
        % \caption{Time Interval 781}
        % \label{TimeInterval781}
     %\end{subfigure}
     %\hfill
     %\begin{subfigure}[b]{0.5\textwidth}
      %   \centering
       %  \includegraphics[width=\textwidth]{1065_optk_new2_new.png}
        % \caption{Time Interval 1065}
         %\label{TimeInterval1065}
     %\end{subfigure}
      %  \caption{Optimal Number of Clusters of Query Results}
       % \label{optimalk}
%\end{figure}

\subsection{Experiment 2}

\begin{figure}
     \centering
     \begin{subfigure}[b]{0.4\textwidth}
         \centering
         \includegraphics[width=\textwidth]{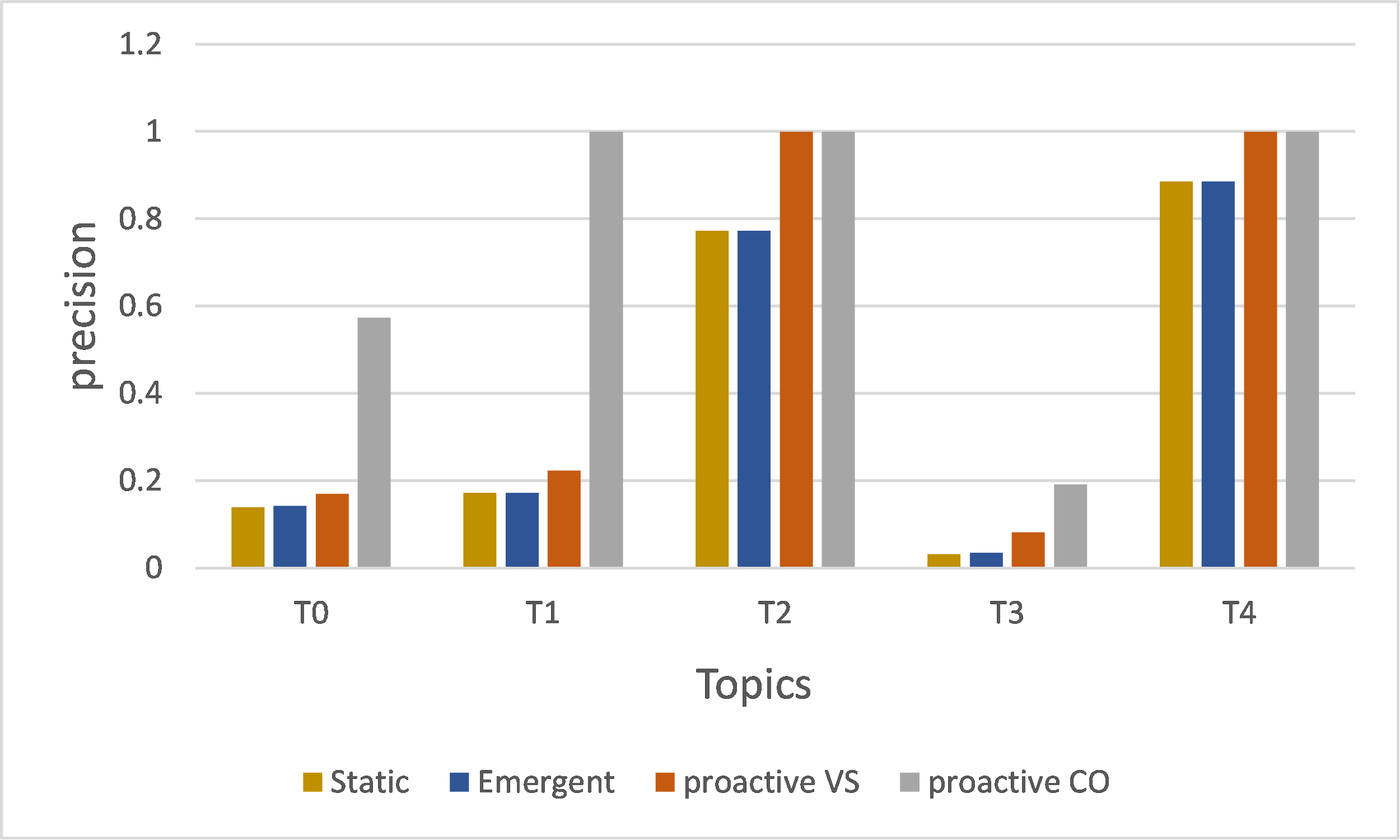}
         \caption{baltimore }
         \label{balt155}
     \end{subfigure}
     \hfill
     \begin{subfigure}[b]{0.4\textwidth}
         \centering
         \includegraphics[width=\textwidth]{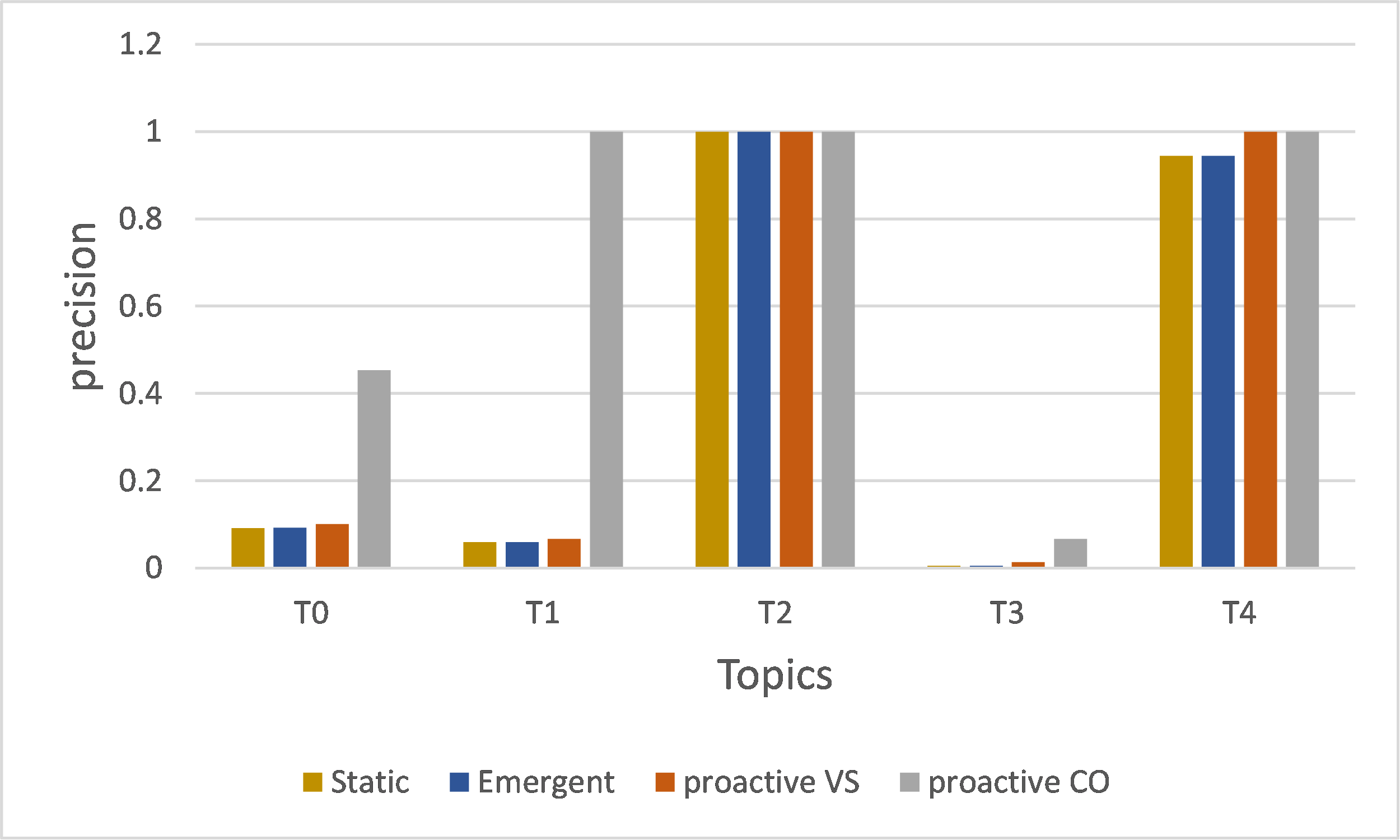}
         \caption{baltimoreprotest}
         \label{baltpro155}
     \end{subfigure}
     \hfill
     \begin{subfigure}[b]{0.4\textwidth}
         \centering
         \includegraphics[width=\textwidth]{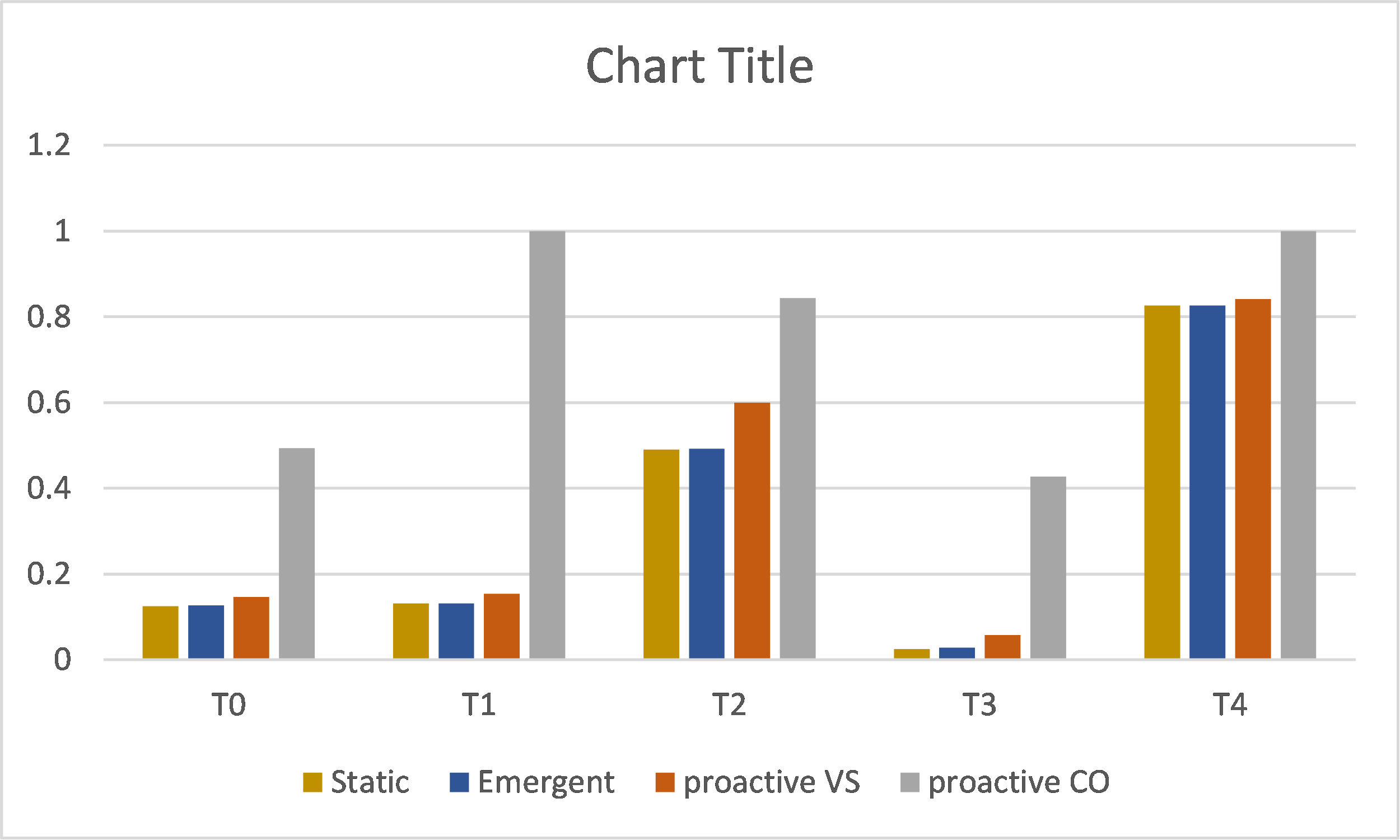}
         \caption{blacklivesmatter}
         \label{blacklive155}
     \end{subfigure}
     \hfill
     \begin{subfigure}[b]{0.4\textwidth}
         \centering
         \includegraphics[width=\textwidth]{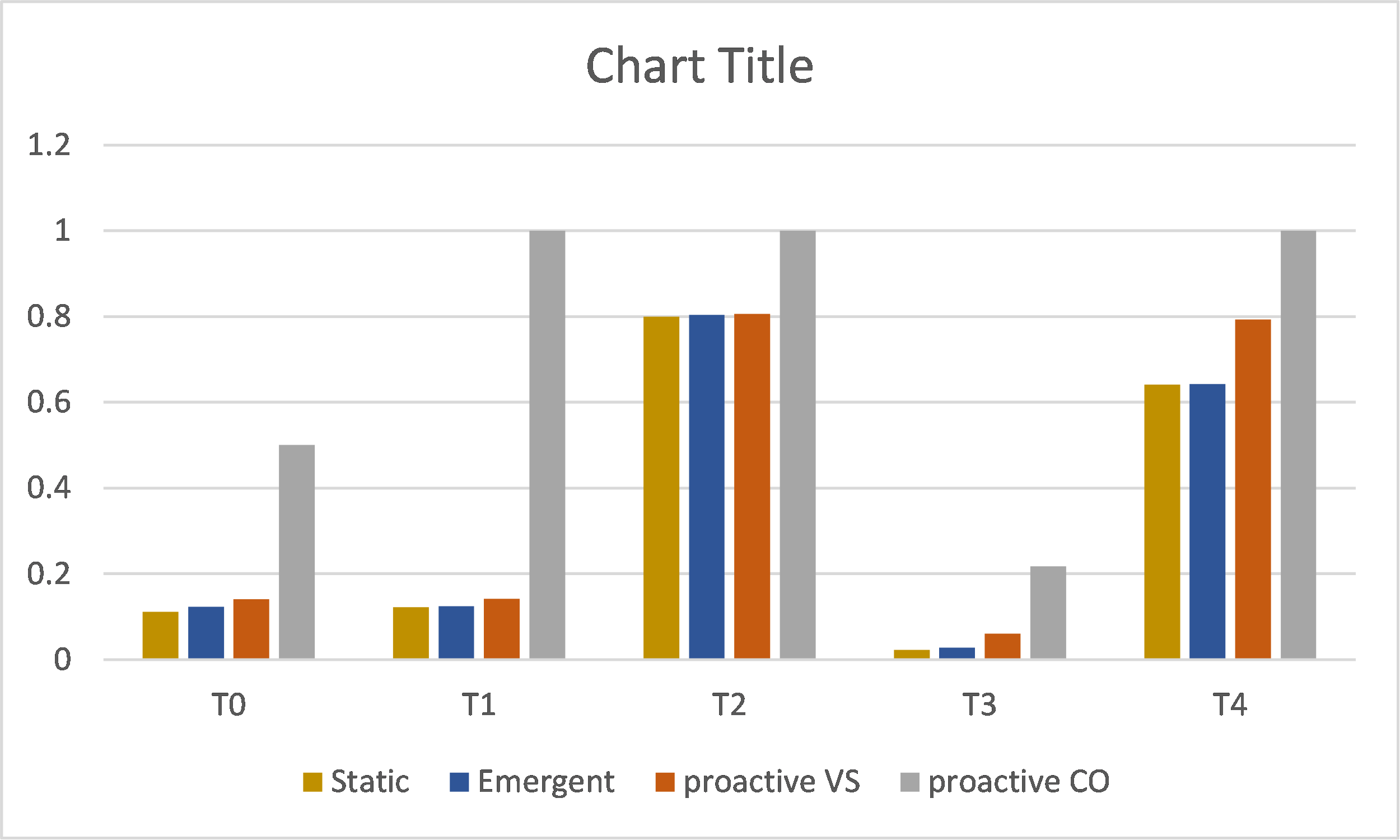}
         \caption{protest}
         \label{protest155}
     \end{subfigure}
        \caption{Prediction the events that happened in time interval 781 from time interval 155 using the highest frequency hashtags}
        \label{pred155}
\end{figure}

\begin{figure}
     \centering
     \begin{subfigure}[b]{0.4\textwidth}
         \centering
         \includegraphics[width=\textwidth]{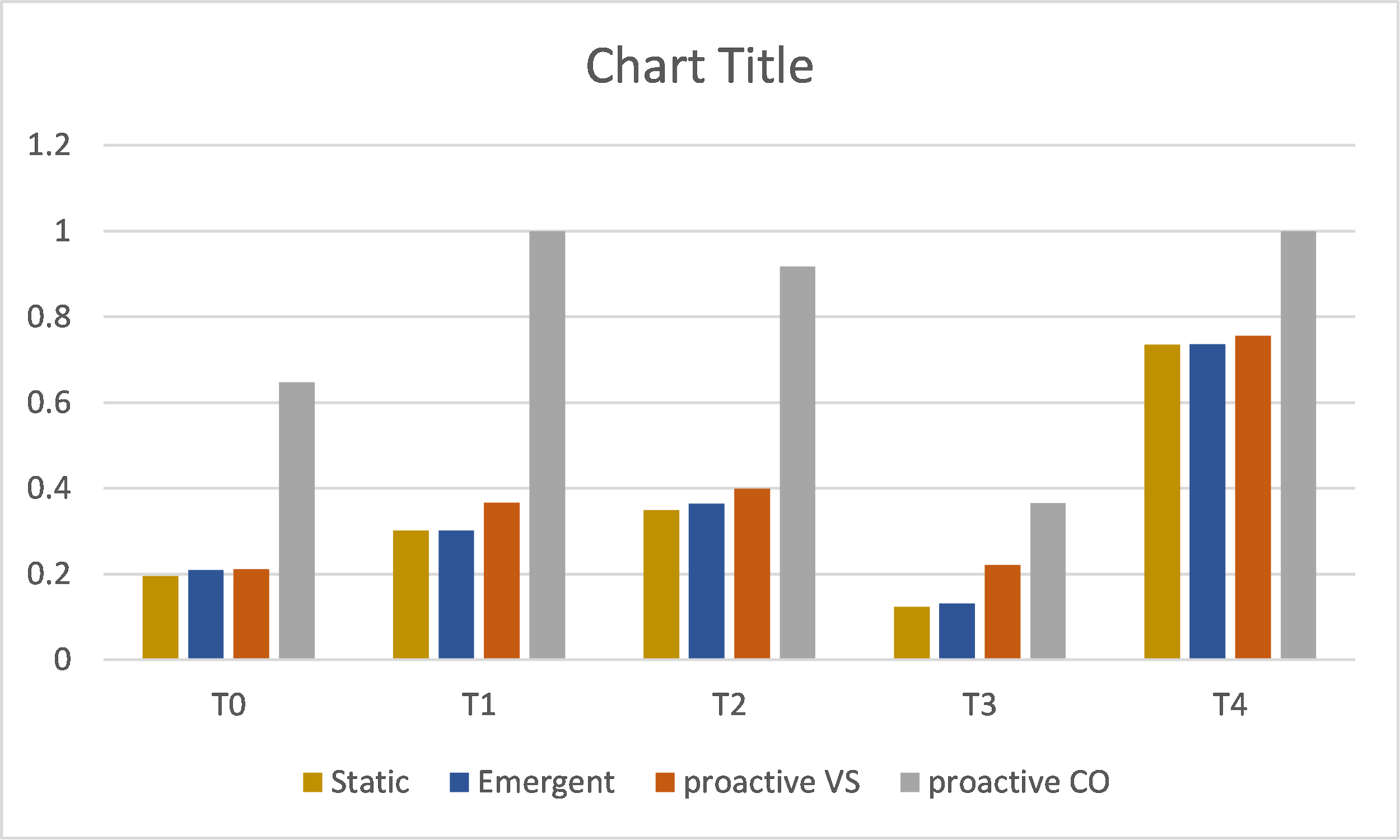}
         \caption{health }
         \label{health155}
     \end{subfigure}
     \hfill
     \begin{subfigure}[b]{0.4\textwidth}
         \centering
         \includegraphics[width=\textwidth]{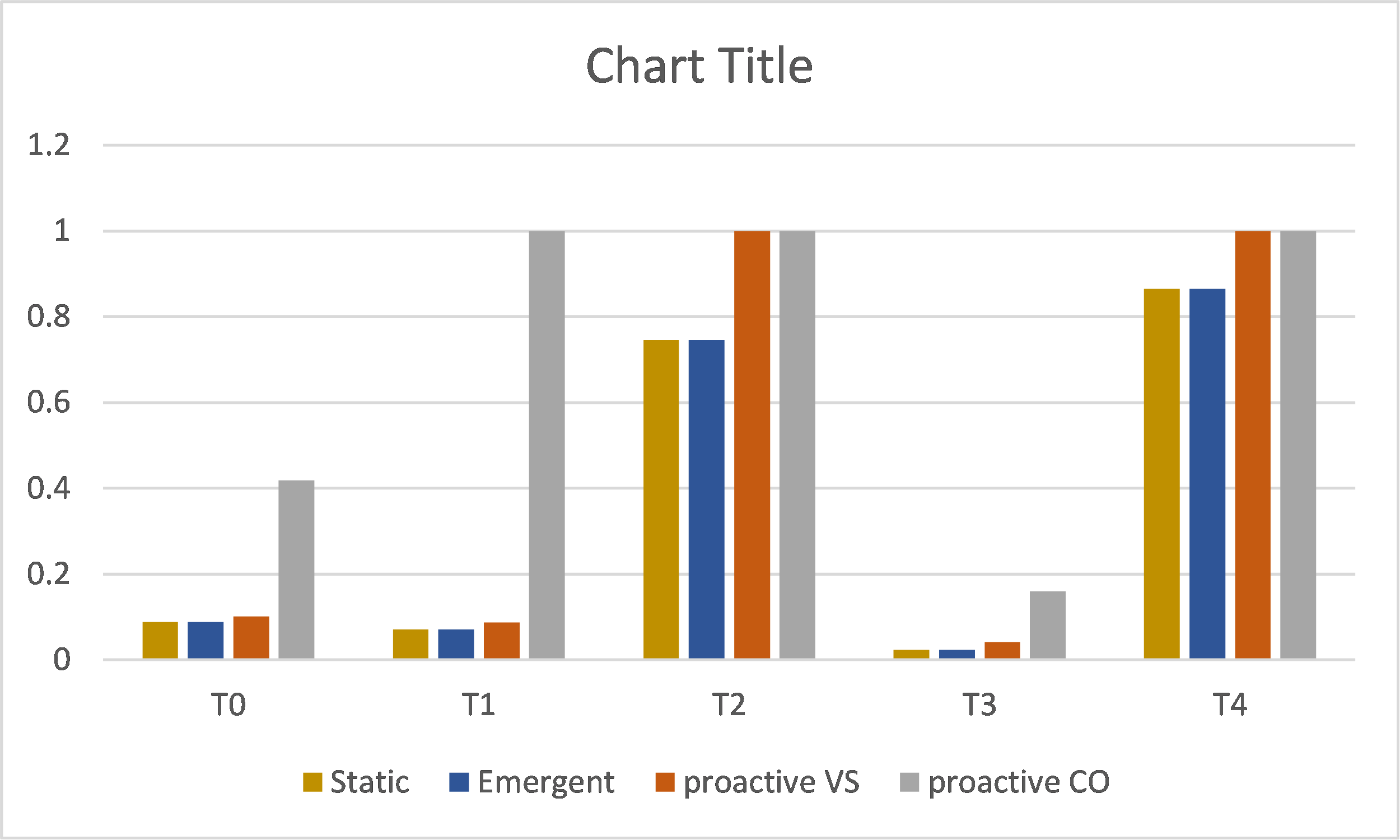}
         \caption{prayforbaltimore}
         \label{prayforbaltimore155}
     \end{subfigure}
     \hfill
     \begin{subfigure}[b]{0.4\textwidth}
         \centering
         \includegraphics[width=\textwidth]{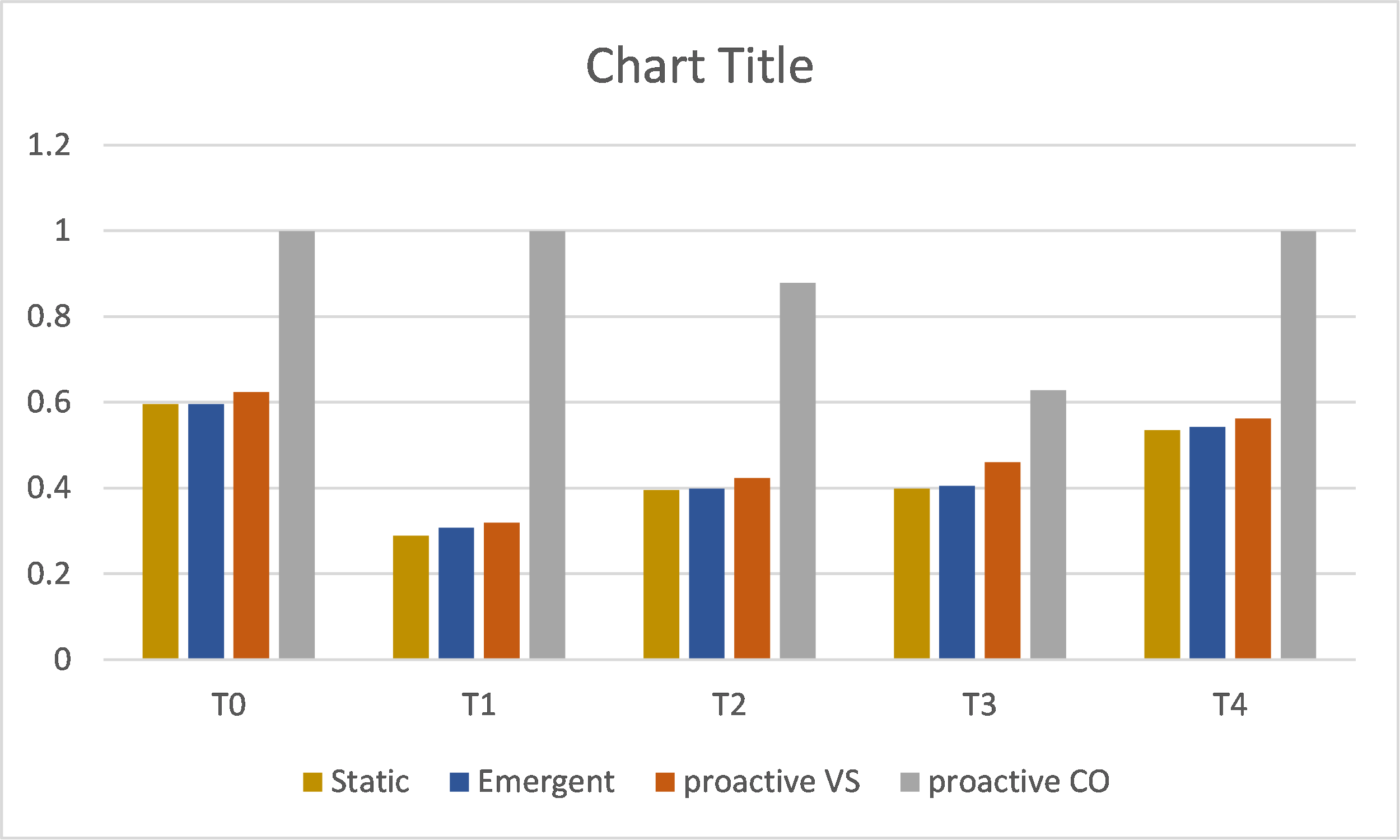}
         \caption{murder}
         \label{murder155}
     \end{subfigure}
     \hfill
     \begin{subfigure}[b]{0.4\textwidth}
         \centering
         \includegraphics[width=\textwidth]{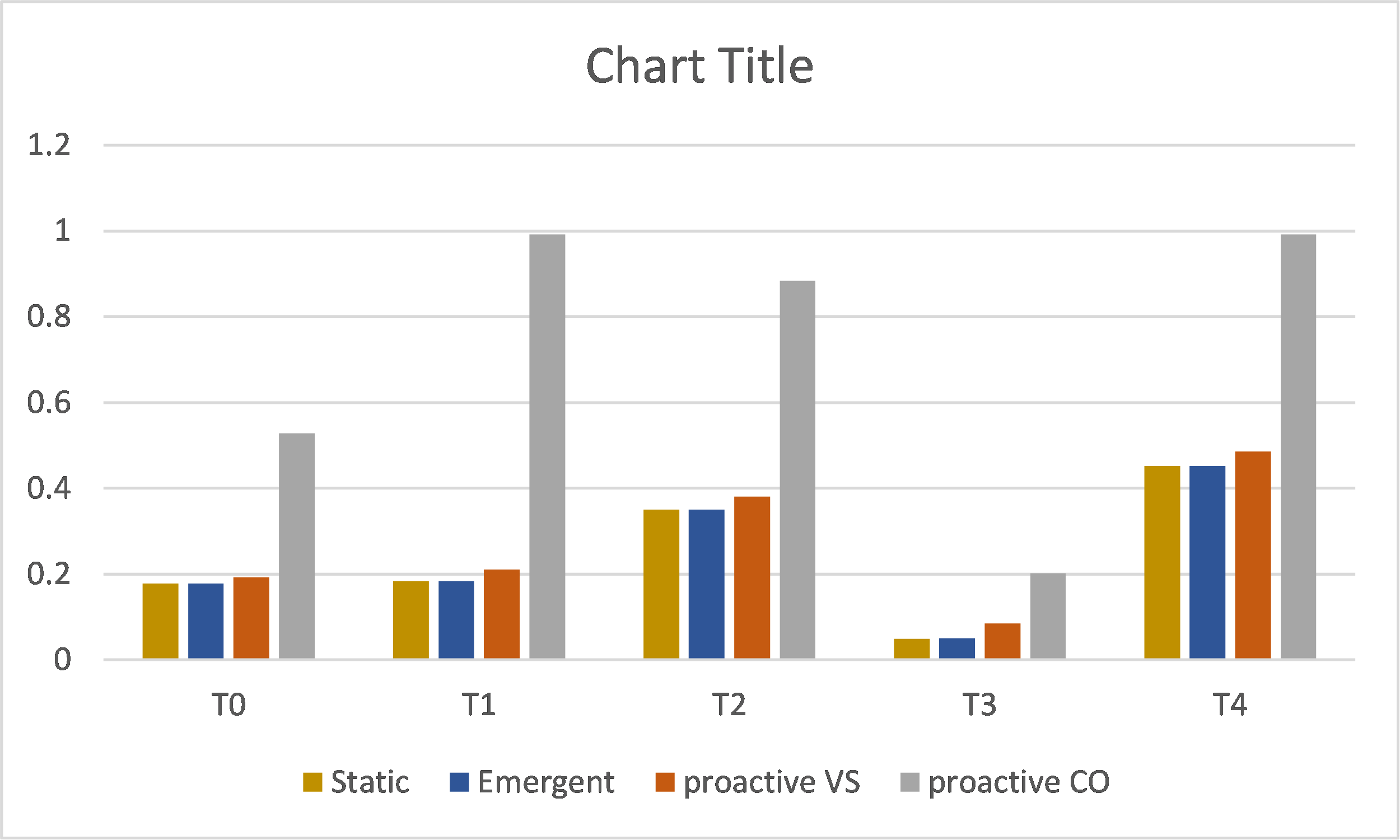}
         \caption{racism}
         \label{racism155}
     \end{subfigure}
        \caption{Prediction the events that happened in time interval 781 from time interval 155 using the lowest frequency hashtags}
        \label{pred155-2}
\end{figure}

In this experiment we test the effectiveness of our proposed methods to predict future emerging events or to predict the conversations from previous time intervals using the precision of a different set of hashtags.  For example, if we are interested in predicting the events that might happen as a result of the death of Frediy Gray. Let us assume that we use the time interval $x$ which has the tweets about this event (death of Frediy Gray) to return the query results (set of tweets) that satisfies the expanded query for that interval, and we will use hashtags from time interval $x+n$ that has tweets about the events happened based on his death. To predict the events that happened in time interval $x+n$ from time interval $x$, we picked a random set of hashtags consists of the highest and lowest frequency hashtags from time interval $x+n$ such as ‘\#protest’ to check if the query results from time interval $x$ contain these hashtags using the precision. We define precision as a fraction of the relevant hashtags. For example, the precision to predict \#protest that appears in time interval  $x+n$  from time interval $x$ is the count of the hashtag “\#protest” in the query results at time interval $x$ divided by the count of the hashtag “\#protest” from the start of the time interval $x$ to the end of primary stream. If the precision is near or equal to one indicates the effectiveness of the method to predict future events as shown in Figures \ref{pred155} and \ref{pred155-2} represents the results of the four query expansion methods in predicting the events that happened in time interval 781 from time interval 155  using highest and lowest hashtags. It is noticed that the precision of our proactive VS and proactive CO is higher than other methods static and emergent for all the topics. For example, as shown in Figure.\ref{pred155} (d), proactive CO predicts the topics about the future event "protest" that appears in topics 1, 2, and 4 which indicates its effectiveness. Additionally, we can say proactive VS and proactive CO can target all the events that appear in streaming data even if it has a few hashtags that describe it. For instance, hashtag “\#health” appears 12 times compared to hashtag “\#baltimore” that appears 1207 in the time interval 781, however, our methods were able to capture the topics that have these events as appear in \ref{pred155-2} (a).

\begin{figure}
     \centering
     \begin{subfigure}[b]{0.4\textwidth}
         \centering
         \includegraphics[width=\textwidth]{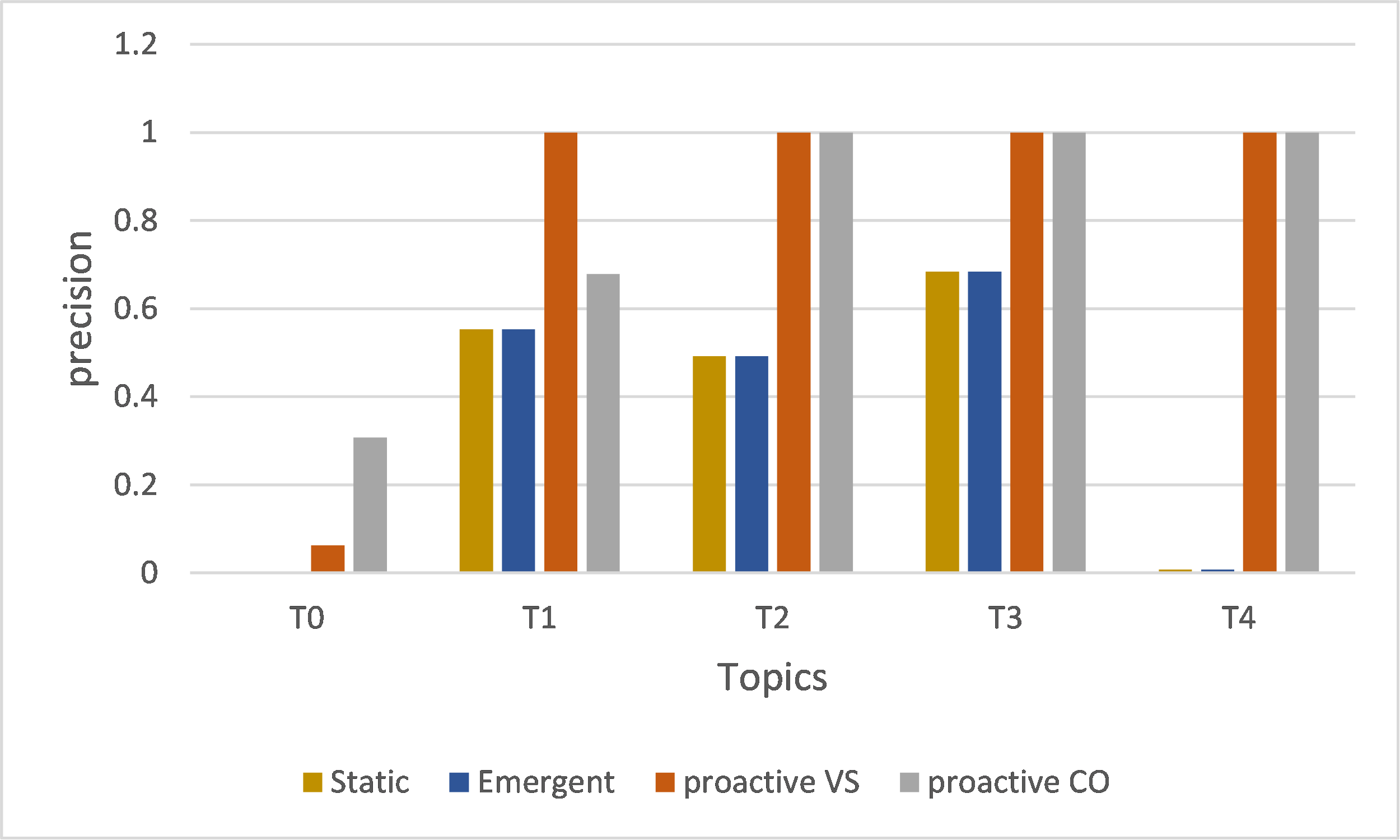}
         \caption{baltimoreuprising }
         \label{baltup781}
     \end{subfigure}
     \hfill
     \begin{subfigure}[b]{0.4\textwidth}
         \centering
         \includegraphics[width=\textwidth]{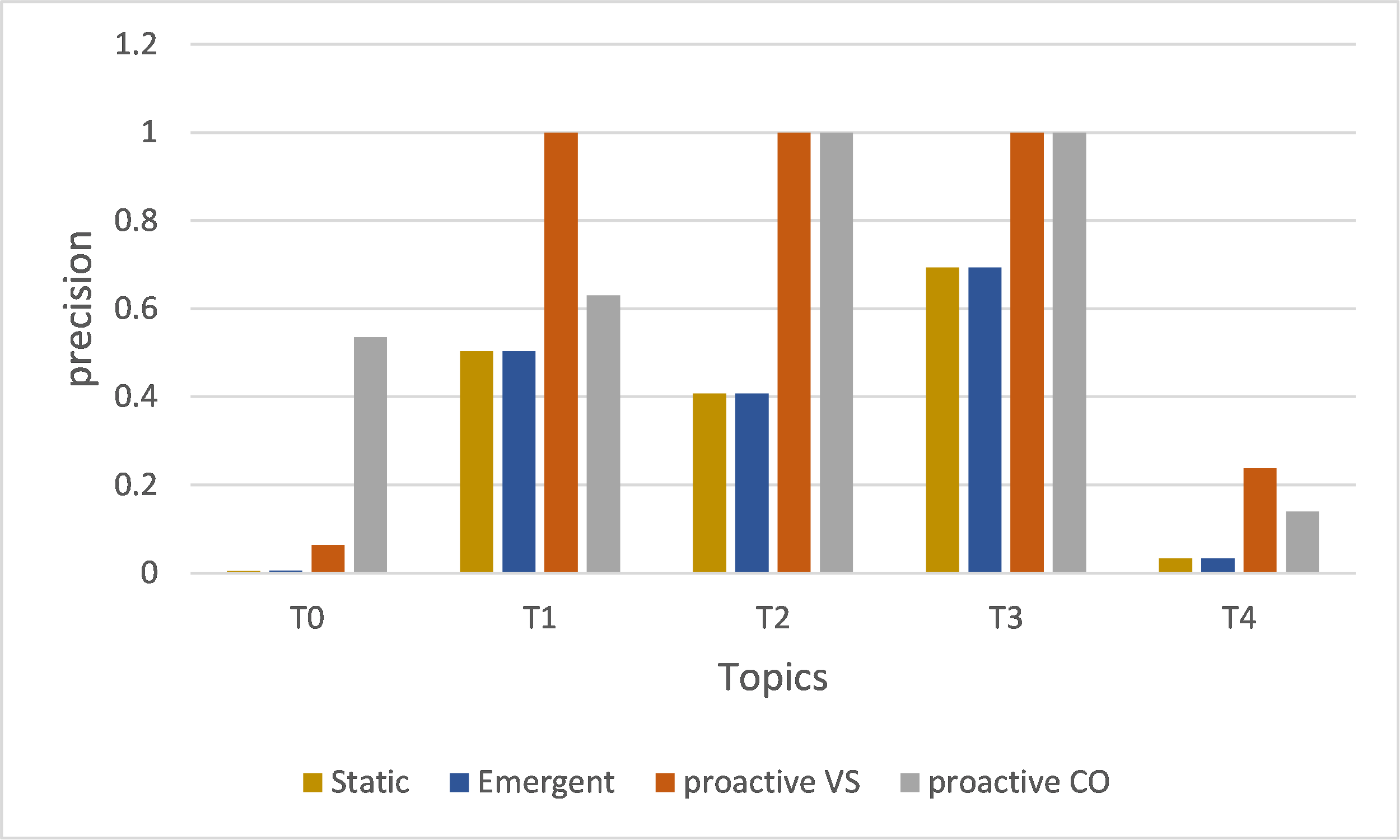}
         \caption{curfew}
         \label{curfew781}
     \end{subfigure}
     \hfill
     \begin{subfigure}[b]{0.4\textwidth}
         \centering
         \includegraphics[width=\textwidth]{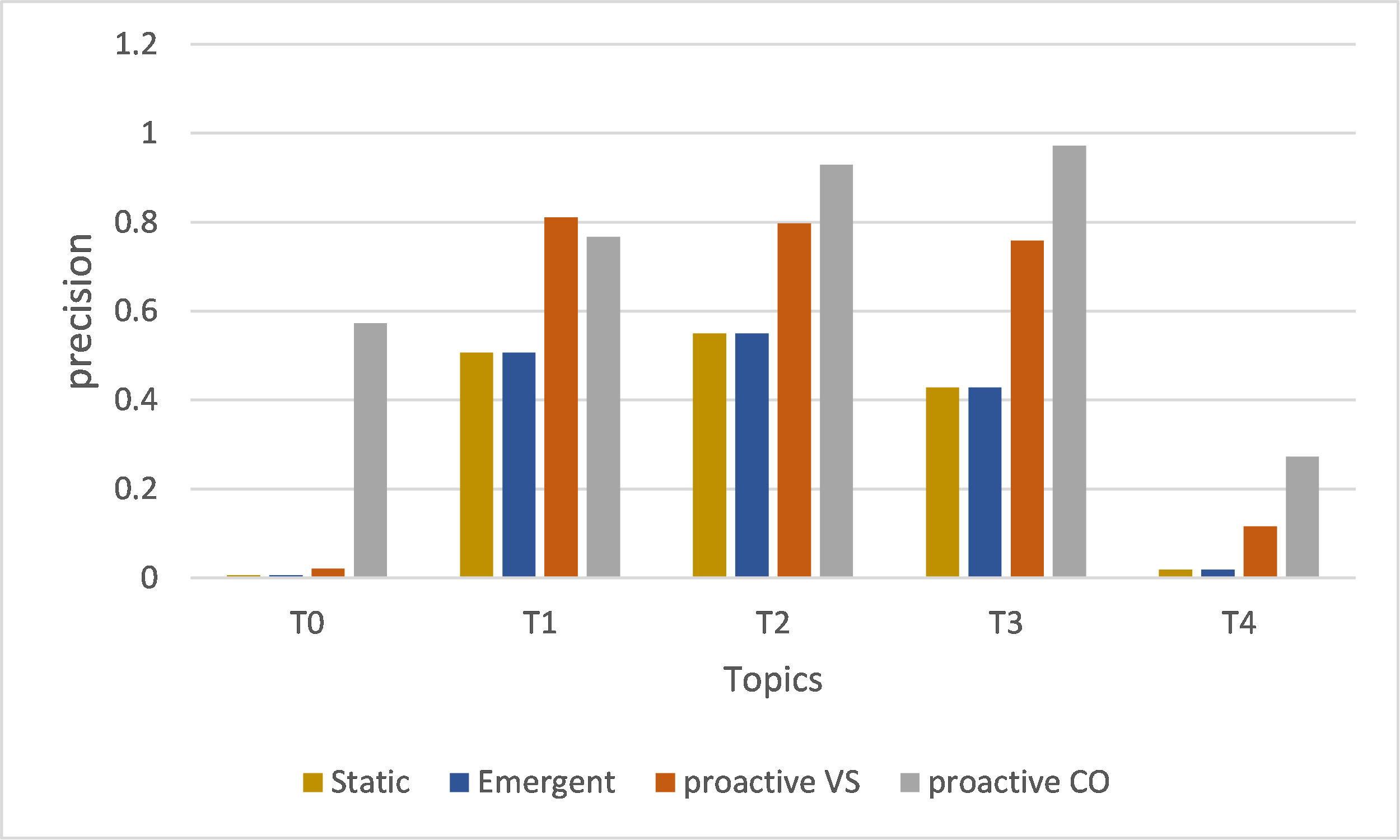}
         \caption{baltimorecurfew}
         \label{baltimorecurfew781}
     \end{subfigure}
     \hfill
     \begin{subfigure}[b]{0.4\textwidth}
         \centering
         \includegraphics[width=\textwidth]{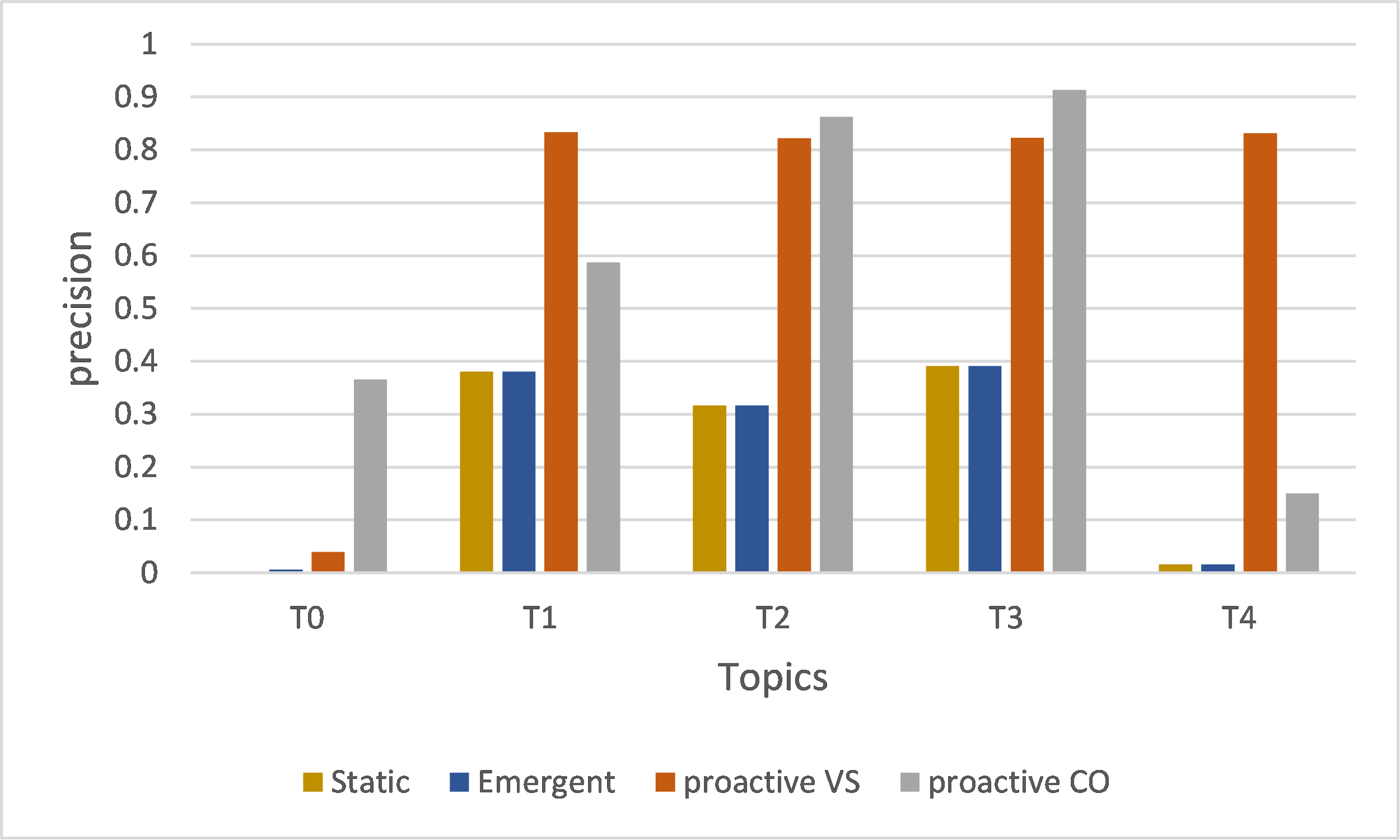}
         \caption{breaking}
         \label{breaking781}
     \end{subfigure}
        \caption{Prediction the events that happened in time interval 1065 from time interval 781 using the highest frequency hashtags}
        \label{pred781}
\end{figure}

\begin{figure}
     \centering
     \begin{subfigure}[b]{0.4\textwidth}
         \centering
         \includegraphics[width=\textwidth]{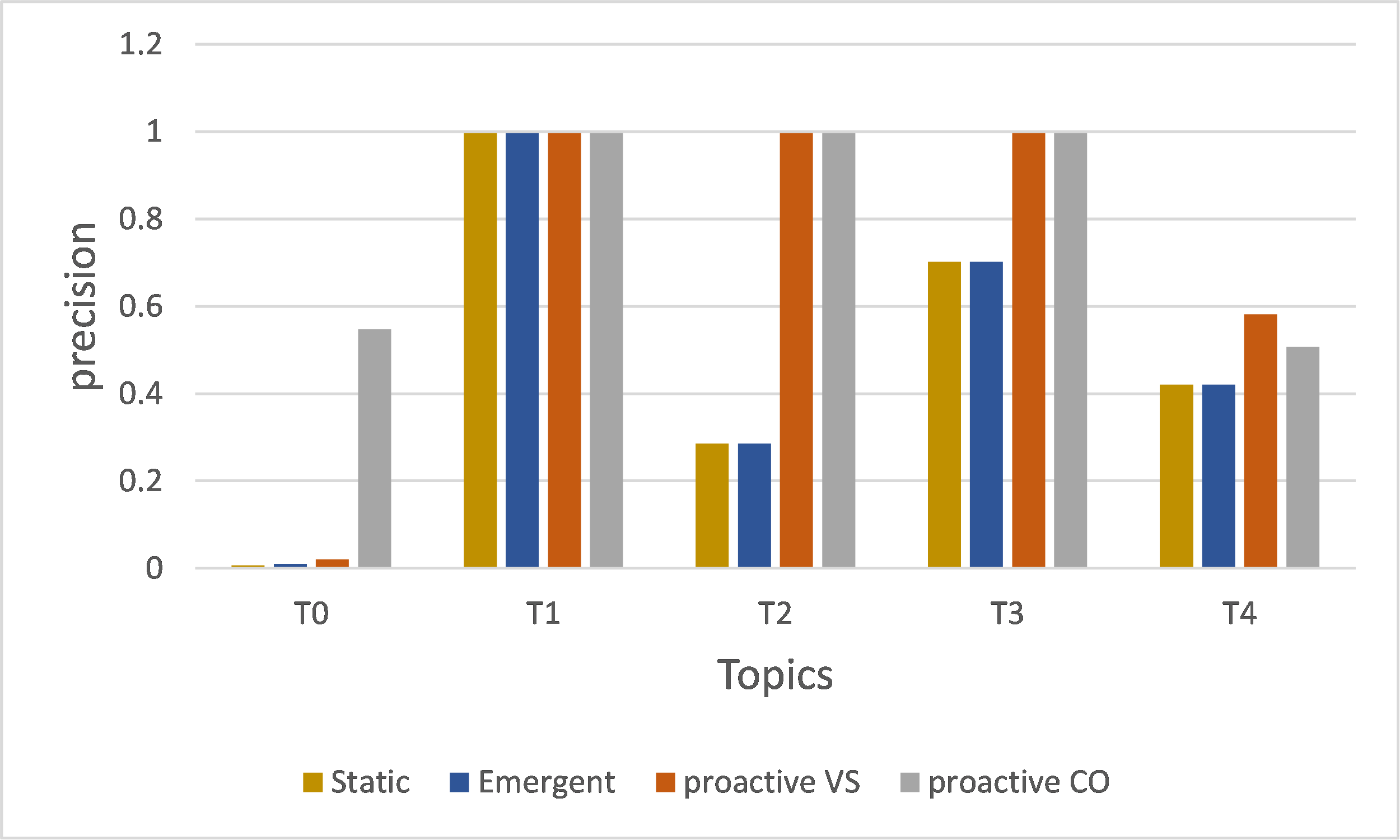}
         \caption{peace }
         \label{peace781}
     \end{subfigure}
     \hfill
     \begin{subfigure}[b]{0.4\textwidth}
         \centering
         \includegraphics[width=\textwidth]{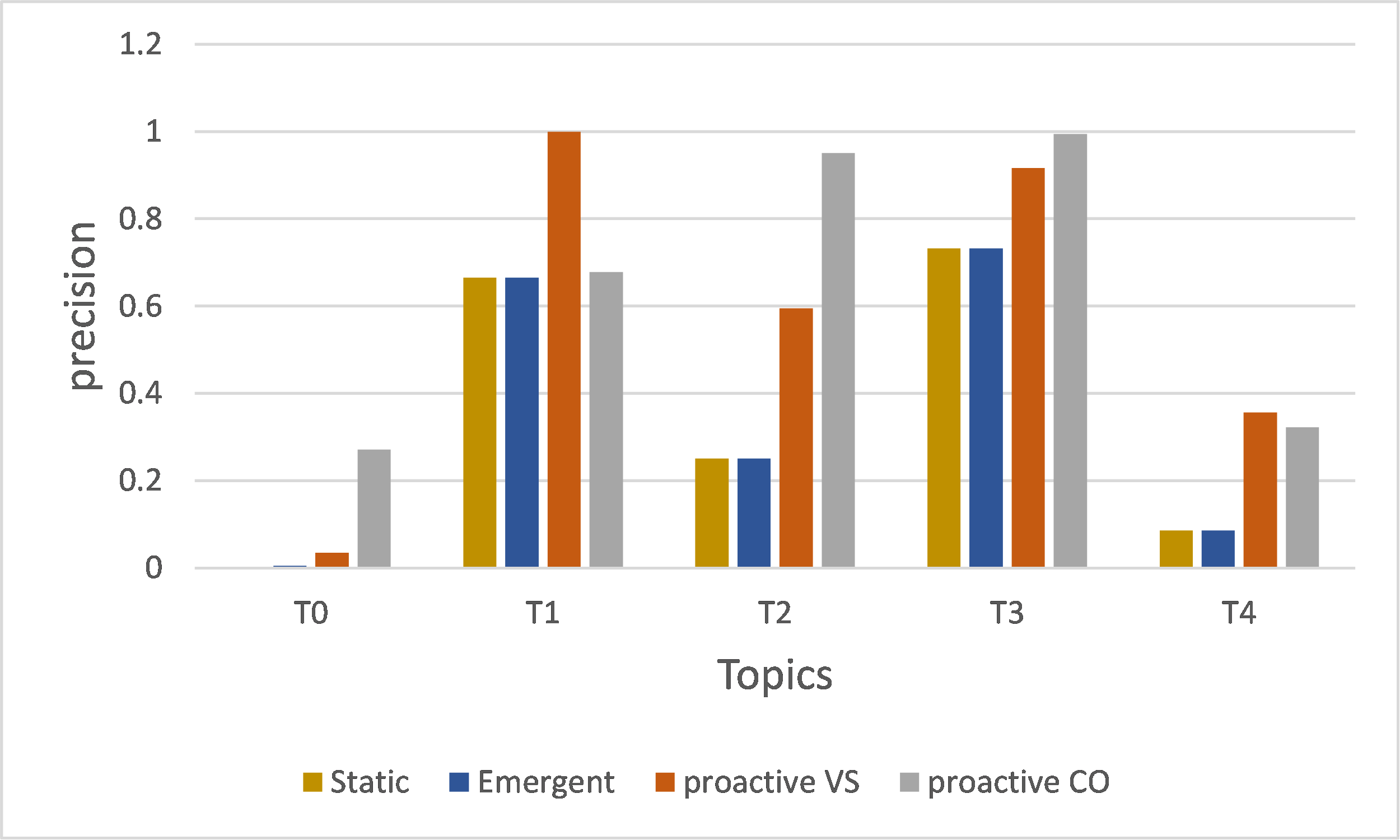}
         \caption{shutdownpalace}
         \label{shutdownpalace781}
     \end{subfigure}
     \hfill
     \begin{subfigure}[b]{0.4\textwidth}
         \centering
         \includegraphics[width=\textwidth]{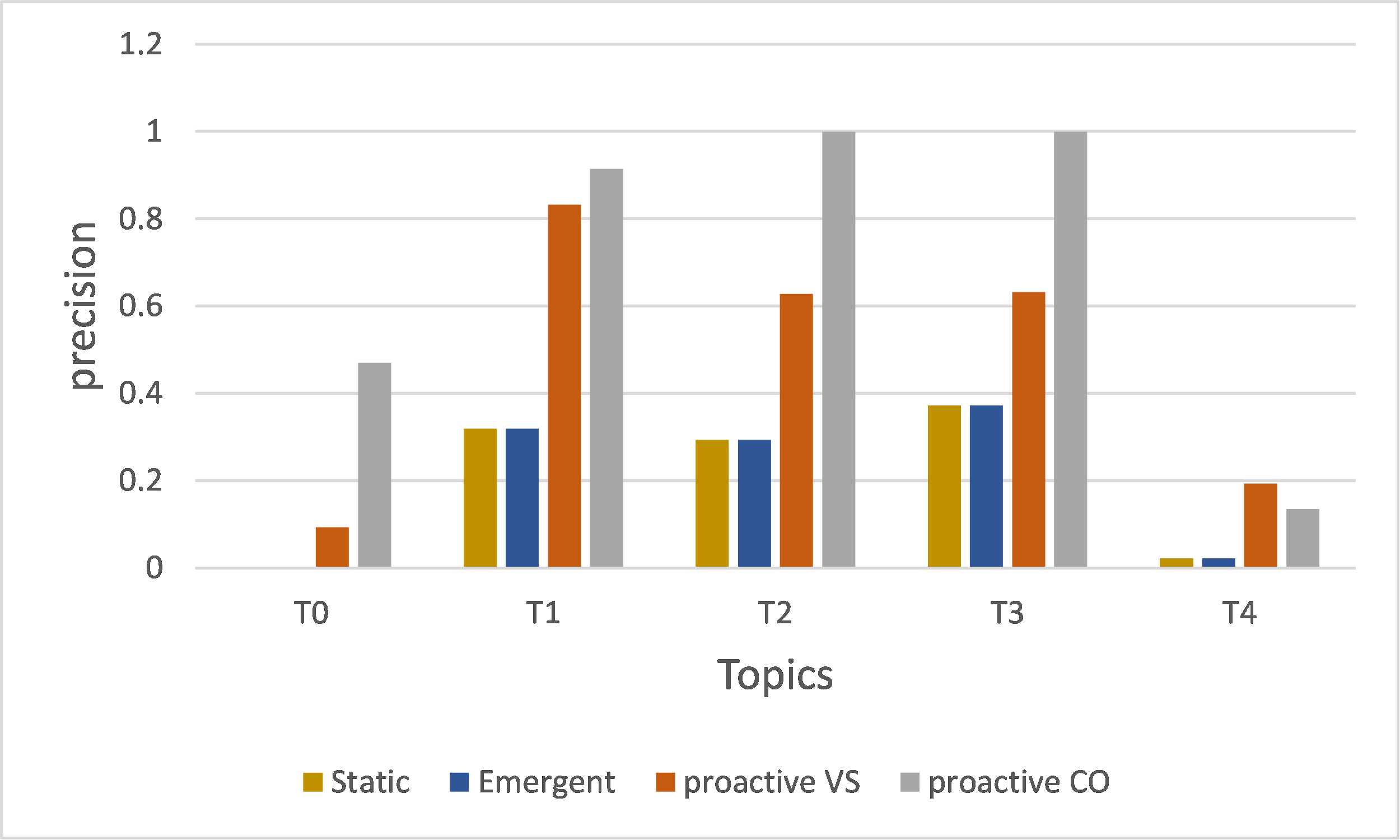}
         \caption{peaceinbaltimore}
         \label{peaceinbaltimore781}
     \end{subfigure}
     \hfill
     \begin{subfigure}[b]{0.4\textwidth}
         \centering
         \includegraphics[width=\textwidth]{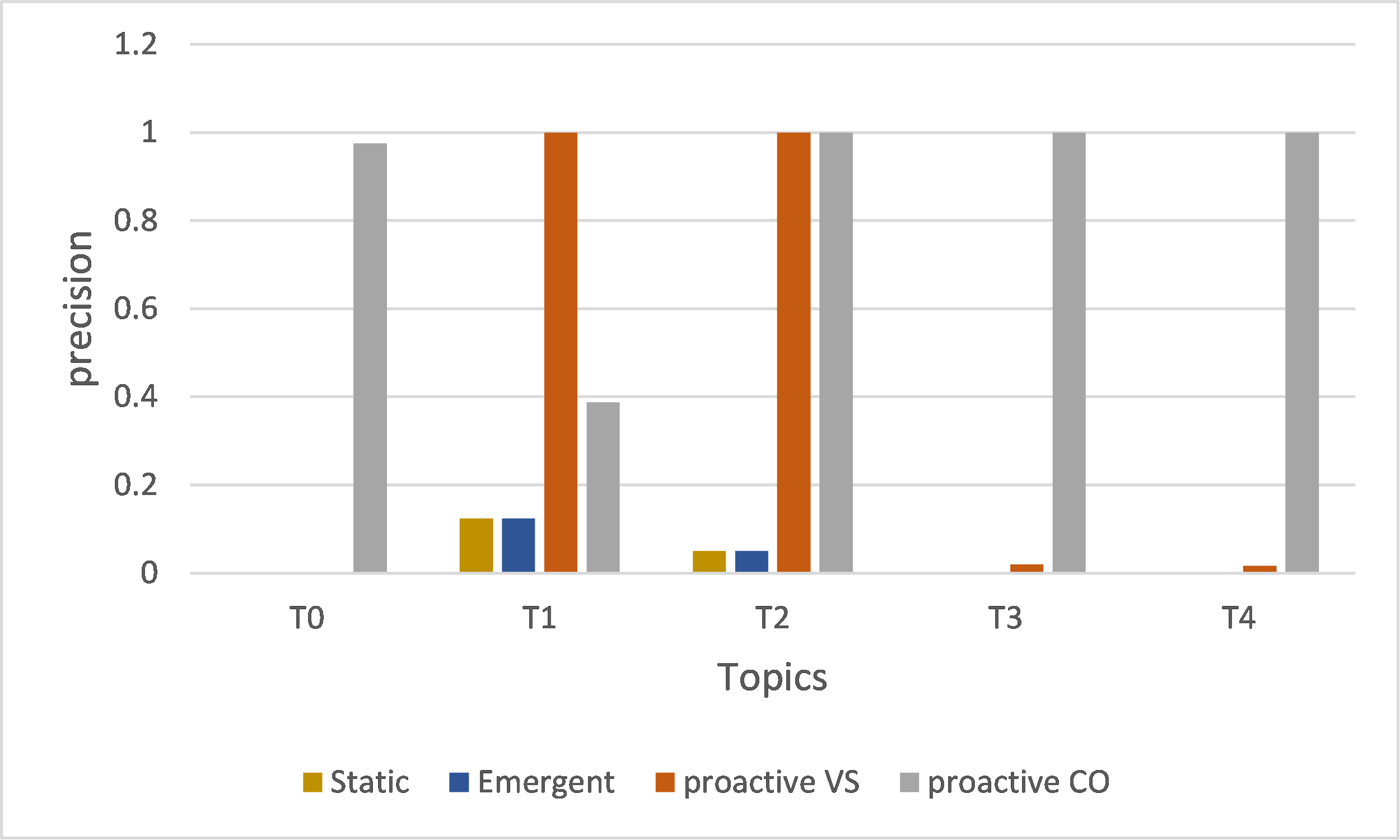}
         \caption{policelivesmatter}
         \label{policelivesmatter781}
     \end{subfigure}
        \caption{Prediction the events that happened in time interval 1065 from interval 781 using the lowest frequency hashtags}
        \label{pred781-2}
\end{figure}

Figures \ref{pred781} and \ref{pred781-2} present the precision for predicting the events that happened in the time interval 1065 from the time interval 781 using the highest  and lowest frequency hashtags. We find that our proactive query expansion methods have  the  same  trend  as  those  observed  in the  two Figures \ref{pred781} and \ref{pred781-2}.

\section{Conclusion}
This study proposed a novel approach for expanding  queries called the proactive query expansion method. This new method depends on adding novel words to the search query from an external data source, where the words are chosen using either nearest neighbor words or highest frequency words to each word that appears in the five LDA topics and DEC words.  Two major experiments were performed: (1) we compared the performance of our proposed query expansion methods:  Proactive VS and Proactive CO (query expansion using LDA, DEC, and external data) with the reference methods (Static: query expansion using LDA) and Emergent (query expansion using LDA and DEC). The performance of the proposed approach is quantified by quality indicators of the streaming data which are the tweet count, hashtag count, and hashtag clustering. (2) We tested the effectiveness of our proposed methods to predict future emerging events or to predict the conversations from previous time intervals using a different set of hashtags. Our experiment performed on 20.5 million tweets (primary stream) covers fifteen days from April 17–May 3, 2015.  For our external source, the external data (secondary stream), we collected news from CNN and New York Times which covers one year before the event happened (2014).  
Generally, the proactive query expansion methods (Proactive VS and Proactive CO) improve the performance of the information retrieval and achieve higher performance compared with Static and Emergent. Additionally, the experiments indicate that our approach can enhance the quality of the results for all the topics. Besides, our proposed methods are more concise comparing to Static and Emergent. Finally, the proposed methods play a key role in enhancing the performance of the search query, which means providing the user with more relative and concise results of interest. As a possible future research direction, we suggest evaluating the proactive query expansion within domains of interest (such as health or social activity) combined with domain specific text preprocessing such as elimination of domain related stop words \cite{alshanik2020accelerating}.

\bibliography{sample}

\end{document}